\documentclass[manuscript,screen]{acmart}

\AtBeginDocument{%
  }

\usepackage{amsmath,amsfonts}
\usepackage{dsfont}
\usepackage{array}
\usepackage[caption=false,font=normalsize,labelfont=sf,textfont=sf]{subfig}
\usepackage{textcomp}
\usepackage{stfloats}
\usepackage{url}
\usepackage{verbatim}
\usepackage{graphicx}
\usepackage{xcolor}
\usepackage{booktabs}
\usepackage{multirow}
\usepackage{adjustbox}
\usepackage{booktabs}
\usepackage{multirow}
\usepackage[normalem]{ulem}
\usepackage{graphicx}
\usepackage{subcaption}
\usepackage{svg}
\usepackage{xcolor}
\usepackage{ulem}
\usepackage{tabularx}
\usepackage{subcaption}
\usepackage[linesnumbered,ruled,vlined]{algorithm2e}

\setcopyright{acmlicensed}
\copyrightyear{2018}
\acmYear{2018}
\acmDOI{XXXXXXX.XXXXXXX}
\acmConference[Conference acronym 'XX]{Make sure to enter the correct
  conference title from your rights confirmation email}{June 03--05,
  2018}{Woodstock, NY}
\acmISBN{978-1-4503-XXXX-X/2018/06}

\begin{document}

\title{CPGRec+: A Balance-oriented Framework for Personalized Video Game Recommendations}


\author{Xiping Li}
\affiliation{%
  \institution{Harbin Institute of Technology}
  \city{Shenzhen}
  \state{Guangdong}
  \country{China}}
\email{lihsiping@gmail.com}

\author{Aier Yang}
\affiliation{%
  \institution{Harbin Institute of Technology}
  \city{Shenzhen}
  \country{China}}
\email{yangaier0920@gmail.com}

\author{Jianghong Ma}
\authornote{Corresponding author}
\affiliation{%
  \institution{Harbin Institute of Technology}
  \city{Shenzhen}
  \country{China}}
\email{majianghong@hit.edu.cn}

\author{Kangzhe Liu}
\affiliation{%
  \institution{Harbin Institute of Technology}
  \city{Shenzhen}
  \country{China}}
\email{kangzheliu@foxmail.com}


\author{Shanshan Feng}
\affiliation{%
    \institution{School of Computer Science, Wuhan University}
  \city{Wuhan}
  \country{China}}
\email{victor_fengss@whu.edu.cn}

\author{Haijun Zhang}
\affiliation{%
  \institution{Harbin Institute of Technology}
  \city{Shenzhen}
  \country{China}}
\email{hjzhang@hit.edu.cn}

\author{Yi Zhao}
\affiliation{%
  \institution{Harbin Institute of Technology}
  \city{Shenzhen}
  \country{China}}
\email{zhao.yi@hit.edu.cn}

\renewcommand{\shortauthors}{Xiping Li et al.}

\begin{abstract}
The rapid expansion of gaming industry requires advanced recommender systems tailored to its dynamic landscape. Existing Graph Neural Network (GNN)-based methods primarily prioritize accuracy over diversity, overlooking their inherent trade-off. To address this, we previously proposed CPGRec, a balance-oriented gaming recommender system. However, CPGRec fails to account for critical disparities in player-game interactions, which carry varying significance in reflecting players' personal preferences and may exacerbate over-smoothness issues inherent in GNN-based models. Moreover, existing approaches underutilize the reasoning capabilities and extensive knowledge of large language models (LLMs) in addressing these limitations.
To bridge this gap, we propose two new modules. First, Preference-informed Edge Reweighting (PER) module assigns signed edge weights to qualitatively distinguish significant player interests and disinterests while then quantitatively measuring preference strength to mitigate over-smoothing in graph convolutions. Second, Preference-informed Representation Generation (PRG) module leverages LLMs to generate contextualized descriptions of games and players by reasoning personal preferences from comparing global and personal interests, thereby refining representations of players and games. Experiments on \textcolor{black}{two Steam datasets} demonstrate CPGRec+'s superior accuracy and diversity over state-of-the-art models. The code is accessible at \url{https://github.com/HsipingLi/CPGRec-Plus}.
\end{abstract}



\received{23 March 2025}
\received[revised]{29 August 2025}
\received[accepted]{29 December 2025}

\maketitle

\section{Introduction}

Recommender systems, pivotal in the customization of digital media content, have gained significant traction across a spectrum of industries, including e-commerce \cite{9492753, app13053347, 9171850, 10247593, KHAN2021107552,jiang2023nah}, news \cite{ULIAN2021115341, SEMENOV2022116478, 9449913, ZHANG2022107922, 9800139,wu2023personalized,newsrec1,newsrec2}, and music \cite{9204829, 9709542, 9546676, 9716820, 9616385,la2022music,musicrec1,musicrec2,musicrec3}. Notably, the gaming sector has witnessed an extraordinary surge in recent years. A compelling example is provided by the data from SteamDB\footnote{https://steamdb.info/stats/releases/}, which indicates a remarkable proliferation of games on the Steam platform: by 2024, the total number of game titles released had reached 14,310, representing an astonishing 33-fold increase over the past decade. This expansive selection of games highlights the substantial potential and the pressing need for the development of recommendation algorithms tailored for the gaming industry. These algorithms are poised to elevate user engagement and bolster financial performance for platforms, developers, and publishers alike, thus creating a symbiotic relationship.

The domain of game recommendation studies has recently attracted considerable academic attention, primarily focusing on developing accuracy-oriented recommender systems to provide tailored game suggestions \cite{ikram2022multimedia,Yang_2022, 9675508, inproceedings, Pérez-Marcos2020,caroux2023presence}. Despite the considerable advancements these models have achieved in enhancing accuracy, they usually neglect the essential diversity of the recommendation lists, raising concerns about the filter bubble effect \cite{wolfowicz2023examining, han2022ssle, knobloch2023algorithmic, michiels2022filter}, where users are confined to a narrow selection of popular games. This limitation may curtail users' exposure to a variety of perspectives and potentially exacerbate pre-existing biases. By integrating diversity into recommendations, the user experience can be significantly enriched, introducing fresh and unexpected choices that broaden horizons and cater to a wider demographic. Against this background, the domain of game recommendation studies is currently facing several key obstacles:

\begin{itemize}
    \item \textbf{Limitations of Accuracy-Focused Methods.} Traditional accuracy-centric game recommendation algorithms \cite{7868073, 9675508, inproceedings, Pérez-Marcos2020} emphasize transformantions from spatial domain but underutilize rich categorical game data. Notably, games with similar genres can originate from various publishers and development studios, which significantly influence user preferences. SCGRec \cite{Yang_2022} leverages category-based connections within game graphs to improve accuracy but relies on single-category associations, which may not always reflect true game similarities, potentially reducing recommendation accuracy.

    \item \textbf{Challenges in Diversity-Focused Methods.} Graph Neural Network (GNN)-based diversity-focused methods have introduced techniques like dynamic neighbor sampling \cite{Zheng_2021,10.1145/3460231.3478845}, neighbor selection \cite{Yang_2023}, and BPR loss modifications \cite{rendle2012bpr} to enhance diversity. However, these approaches face two challenges: (1) complexity risks and suboptimal use of item details in neighbor modeling, and (2) the challenge of long-tail distribution in the spread of messages. Neighbor sampling strategies can become overly intricate or sparse, neglecting key item features like categories. Additionally, most methods ignore item popularity, disproportionately favoring popular items and limiting exposure to niche or long-tail games.

    \item \textbf{Balancing Accuracy and Diversity.} The trade-off between accuracy and diversity is a persistent challenge, as these objectives often conflict. Accuracy-focused methods like BPR loss \cite{Yang_2022, wang2019neural, tang2023ranking} refine models using negative examples, while diversity-oriented methods employ loss rescaling or likelihood adjustments \cite{Zheng_2021,Yang_2023,ying2018graph,cheng2017learning}. However, these approaches predominantly favor one objective, grappling with the achievement of a harmonious accuracy-diversity balance.
\end{itemize}

In prior work \cite{cpgrec}, we have addressed the outlined challenges by introducing a novel model, termed Category-based and Popularity-guided Video Game Recommendation (CPGRec). As part of that study, we explored a strategy for constructing a graph structure by explicitly encoding the semantic information of game categories and popularity into the graph. Particularly, (1) when it comes to accuracy, based on the local smoothness of GNNs, CPGRec retains links exclusively between games that share a strong similarity in their categories, emphasizing the rationale for inferring content similarity from category similarity. (2) Regarding diversity, CPGRec allows information to flow freely between games across different categories, breaking the representation clustering of same-category games in high-dimensional space. Furthermore, by limiting the influence of popular game nodes while emphasizing their propagation capability, CPGRec facilitates the widespread dissemination of feature information for long-tail games. (3) Overall, regarding the trade-off between accuracy and diversity, CPGRec infers the relationship between representations and popularity information based on the rating scores of negative samples extracted from the BPR loss, allowing for the reweighting of BPR loss. Experimental results demonstrate that CPGRec outperforms baseline models in both accuracy and diversity.

\textcolor{black}{Despite its success, CPGRec neglects the disparities among observed historical player-game interactions, \textit{i.e.}, it assumes that all these interactions reflect the players' interest equally. However, in real-world scenarios, it is more reasonable to acknowledge the disparities of historical interactions, as they may have different importance in reflecting the player's personal preferences. Intuitively, a player's personal preferences are more pronounced in their historical interactions that significantly differ from those of the broader player base, rather than in interactions that show indistinguishable interest from global interests, considering both positive (interest) and negative (disinterest) aspects (which is analyzed in detail in Section \ref{section: disparity}). Based on this, two points should be highlighted: (1) Considering CPGRec's employment of graph convolutional layers with non-interaction-specific edge weights in bipartite graphs, ignoring these disparities in interactions may cause the smoothness of GCNs (\textit{i.e.,} low-pass filtering in the spectral domain, which leads the adjacent nodes' representations to become similar gradually) to become a trigger, leading to suboptimal modeling of players and games, as well as the resulting representations. (2) Furthermore, although the effectiveness of Large Language Models (LLMs) has been demonstrated in existing works, their great potential in filling the above gap, such as their rich knowledge about the game and powerful reasoning ability for understanding the above disparities of player-game interactions, has not yet been explored.}

To address these problems, we newly propose two modules in this study. They both incorporate both the dwelling time and average ratings, which respectively reflect the personal interest and public interests, to refine the representation of player-game interactions.
\textbf{First}, we design a module, namely \textbf{Preference-informed Edge Reweighting (PER)}, an edge reweighting method with sign design to address the above-mentioned over-smoothing risk arising from the neglect. Both the sign and edge weight are proposed to model the personal preferences indicated by historical interactions, where the sign qualitatively reflects whether significant interest or disinterest is indicated in each interaction, while the edge weight further quantitatively measures the extent of significant preference. Specifically, a two-level approach is adopted: (1) \textbf{Fisher Distribution-based Sign Decision}: Using the Fisher statistic and Fisher distribution, we compare personal and global interests, uncovering both interests and disinterests, which inform the directional flow of messages on each edge, \textit{i.e.}, the sign of messages. (2) \textbf{Information Content-based Volume Evaluation}: We introduce the concept of ``information content'' \cite{info_content, info1, info2, info3} to quantify the significance of each interaction, thereby determining the magnitude of propagated messages. 
\textbf{Second}, inspired by the outstanding performance of existing recommender systems that integrate LLMs, CPGRec+ further proposes a \textbf{Preference-informed Representation Generation (PRG)} module. PRG leverages the broad knowledge as well as powerful reasoning capabilities to generate game descriptions and player descriptions as supplementary representations to polish game representation while further capturing the personal preferences of players. Particularly, PRG operates in two steps: 
(1) \textbf{Rating-informed Game Description Generation}: This step leverages the broad knowledge of LLMs by explicitly instructing them to reason from the average rating of each game, which is highlighted by our designed game prompt. This ensures that the generated descriptions effectively capture and reflect the global interest of the general players toward the game. 
(2) \textbf{Preference-informed Player Description Generation}:  This step utilizes the reasoning capabilities of LLMs, explicitly instructing them to infer player preferences by comparing personal interests with global interests. 
Jointly, these two preference-informed modules prioritize interactions that better reflect player interest to solve these aforementioned problems, thereby improving the accuracy of our enhanced model, CPGRec+.
Experimental results confirm that CPGRec+ outperforms CPGRec and remains competitive with state-of-the-art benchmarks.

Building upon our previous work \cite{cpgrec}, this study presents the following contributions:
\begin{itemize}
    \item We extend CPGRec by newly incorporating both Preference-informed Edge Reweighting (PER) and \textcolor{black}{Preference-informed Representation Generation (PRG)}, enabling finer-grained modeling for player-game historical interactions and personalized preferences. This integration empowers the system’s capability of delivering more tailored and satisfactory recommendations.
    
    \item We propose a methodology to map video-game-related distributions to standard normal distributions using the Box-Cox transformation, which facilitates deeper theoretical analysis of personal preferences embedded within historical interactions and is experimentally validated through hypothesis testing, providing a robust foundation for future game recommendation research.

    \item \textcolor{black}{We propose a novel approach to enhance recommender systems by leveraging LLMs to generate game descriptions based on global interests, and generate player descriptions by reasoning individual preferences from comparing personal interest and global interest, subsequently embedding these descriptions to refine player and game representations for improved recommendation quality.}
    
    \item Experimental results highlight the superior accuracy of the enhanced model, CPGRec+, while maintaining the diversity strengths of CPGRec, thereby attaining an ideal equilibrium for the trade-off task.
\end{itemize}

\section{RELATED WORK}
\subsection{Video Game Recommendation}

The escalating growth of the video game industry has drawn significant scholarly interest in the field of video game recommendation algorithms. Initial scholarly investigations primarily focused on approaches grounded in Collaborative Filtering (CF) and Content-Based Filtering (CBF). Anwar \textit{et al.} presented a CF strategy that addresses the distinct profiles of games and users for the purpose of personalized game endorsements \cite{7868073}. BharathiPriya \textit{et al.} expanded on this by merging CF and CBF to deduce implicit hierarchies, with a focus on elements such as the duration of play \cite{9675508}. P\'{e}rez-Marcos \textit{et al.}, in a related approach, proposed a hybrid recommendation framework for gaming, drawing on methodologies from the music sector to optimize the usage of playtime \cite{Pérez-Marcos2020}.

Recently, there has been a discernible trend towards exploring innovative deep learning methodologies for game recommendation studies. These innovative strategies have consistently surpassed conventional methods, which are largely based on implicit user feedback. A case in point is the comprehensive study by Cheuque \textit{et al.}, which evaluated the traditional Factorization Machine, DeepNN and hybrid DeepFM \cite{inproceedings}.

Their outcomes, notably when utilized with data from the Steam platform, indicated that DeepNN and DeepFM models outperformed the FM model. The integration of GNNs into game recommender systems has also gained considerable traction. Graph Neural Networks (GNNs) excel at capturing intricate connections and dependencies in graph-based data \cite{kipf2017semisupervised, hamilton2018inductive, wei, xie, sun, wen, ying2018graph, gnnrec_1, gnnrec_2, gnnrec_3, gnnrec_4, gnnrec_5, gnnrec_6}, positioning them as a strong choice for advancing recommendation systems in the gaming sector. Yang \textit{et al.} introduced SCGRec, a novel system that uses a GNN architecture to incorporate game context and social interactions \cite{Yang_2022}.

\textbf{Discussion.} CPGRec+ apart from prior research in two key aspects: (1) Unlike traditional game recommendation models overlook the disparities of player-game historical interactions, CPGRec+ deeply mines the rich semantics of interaction-level to provide more accurate recommendations. Building on our earlier work \cite{cpgrec}, it integrates inter-category linkages within the graph structure and employs the Fisher distribution and entropy metrics to enhance player modeling by analyzing personal preferences from historical interactions. \textcolor{black}{Besides, CPGRec+ further leverages the extensive knowledge and advanced reasoning capabilities of LLMs to refine game and player representations, incorporating global interests and conducting a comparative analysis between individual and global interests, respectively.} (2) In contrast to prior studies that often overlook diversity, leading to homogeneous recommendations, CPGRec+ incorporates both accuracy and diversity, ensuring a more balanced and varied recommendation output.

\subsection{Diversified Recommendation}

Pioneers like Ziegler \textit{et al.} \cite{10.1145/1060745.1060754} initially brought the concept of diversity into the domain of recommender systems. This sparked extensive scholarly exploration into diversifying recommendation algorithms. Yin \textit{et al.} \cite{Yin_2023} delved into diversity challenges within session-based recommender systems with a thorough study. Liang \textit{et al.} \cite{liang2021enhancing}'s EDUA employs a dual-path network with adaptive balancing and dual-metric learning to boost both the accuracy and diversity of recommendations. Zheng \textit{et al.}'s DGCN, a strategy based on GCN \cite{kipf2017semisupervised}, boosts diversity by selecting neighbors, reweighting samples, as well as the employment of adversarial learning \cite{Zheng_2021}. Ye \textit{et al.} \cite{10.1145/3460231.3478845}'s DDGraph employs a diversifying selector to update the user-item graph, ensuring neighbor diversity. Yang \textit{et al.} \cite{Yang_2023}'s DGRec uses a submodular function for diverse neighbor selection and employs multi-layer attention and adjusts loss weights to glean more detailed insights from higher-order neighbors. The EXPLORE framework \cite{explore} presents a probabilistic user-behavior model that maximizes recommendation diversity while maintaining relevance, balancing user engagement and exploration.

\textbf{Discussion.} Collectively, the above studies advance the incorporation of diversity in recommender systems. However, most of these works sacrifice accuracy for diversity, use complex dynamic methods inefficiently, and inadequately leverage game category information for neighbor selection. To this end, our recently proposed CPGRec boosts recommendation accuracy by tightly linking games, enhances diversity by connecting varied game categories and leveraging popular games for long-tail visibility, and balances both through a novel negative-sample reweighting approach. 


\textcolor{black}{
\subsection{LLM for Recommendation}
Recent research has explored the application of LLMs to enhance recommendation systems in various ways \cite{llmrec2}.  A-LLMRec proposes an efficient LLM-based recommender system that leverages collaborative knowledge from pre-trained CF-RecSys to excel in both cold and warm scenarios \cite{A-LLMRec}. CoLLM introduces an innovative LLMRec approach that explicitly integrates collaborative information by encoding it from traditional collaborative models and aligning it with the LLM's input text token space \cite{CoLLM}. CUP addresses the challenge of sparse user interactions by using Transformer-based representation learning and selecting informative cues from review texts to construct concise user profiles for recommendations \cite{CUP}. KAR proposes an open-world knowledge-augmented recommendation framework that acquires reasoning knowledge on user preferences and factual knowledge on items from LLMs \cite{KAR}. RLMRec introduces a model-agnostic framework that enhances existing recommenders with LLM-empowered representation learning to capture intricate semantic aspects and address challenges like scalability and text-only reliance \cite{RLMRec}. SAGCN harnesses the sentiment analysis capabilities of LLMs to uncover semantic aspect-aware interactions from user reviews, enhancing the accuracy and interpretability of conventional recommendation methods \cite{SAGCN}.
}

\textbf{Discussion.} Although these latest methods have attempted to leverage the rich knowledge of LLM to generate rich descriptions for users or items, or to encode collaborative filtering signals into representations, they still have some key deficiencies. (1) Existing LLM-based recommender systems often overlook the importance of average ratings, which reflect the preferences of a large number of players, in the process of generating detailed game descriptions. \textcolor{black}{This omission may lead to game descriptions that fail to reflect the aspects most relevant to players' interests. While a game's description may accurately portray its content and initially appear to align with a player's preferences—prompting them to select it—its actual content may fall short of expectations, as indicated by a persistently low average rating. In such cases, despite a superficial alignment based on textual description alone, the game should not be recommended, as the collective feedback from the broader player community reveals a fundamental mismatch.
}
(2) Existing works fail to fully harness the powerful reasoning capabilities of LLMs to extract individual preferences from historical interaction data, particularly by using dwelling time and average ratings as key indicators. These two factors, when combined, provide valuable insights into a player’s unique interests, as discussed in Section \ref{section: disparity}.
\textcolor{black}{To overcome these limitations, CPGRec+ employs large language models (LLMs) to integrate average ratings and dwelling time into the prompting process, enabling the generation of refined game and player descriptions. By explicitly capturing both global player preferences and individual interests, this approach enhances the alignment between game content and user experience, leading to more accurate and personalized recommendations.}

\section{Investigating GNN-based Video Game Recommendation}
\subsection{Preliminaries}
\subsubsection{\textbf{DEFINITION(Game Graphs with Raw Connections)}} 
\label{sec: define raw graph}
In this study, we explore the interconnections among video games by categorizing them based on three fundamental types of categories: genre, developer, and publisher. This framework enables the establishment of raw connections between games. Specifically, in a game graph with genre-based raw connections, a connection forms when two games share overlapping genres; similar approaches define connections based on shared developer and publisher affiliations. By identifying games linked through shared genre (g), developer (d), and publisher (p) characteristics, we create three separate game networks based on these categories, denoted as $\mathcal{G}^{g}$, $\mathcal{G}^{d}$, and $\mathcal{G}^{p}$.

\subsubsection{\textbf{DEFINITION(Player-Game Bipartite Graph)}} In game recommendations, we define a group of players as $\mathcal{U} = \{ u_{1}, u_{2}, \ldots, u_{|\mathcal{U}|} \}$ and a set of games as $\mathcal{I} = \{i_{1}, i_{2}, \ldots, i_{|\mathcal{I}|}\}$. The subset $\mathcal{I}{(u)}\subset \mathcal{I}$ signifies the games that the player \textit{u} has played. To model these historical interactions between players and games, we construct a player-game bipartite graph, represented as $\mathcal{G} = (\mathcal{V},\mathcal{E})$, where $\mathcal{V} = \mathcal{U} \cup \mathcal{I}$. An edge is established between player  \textit{u} and game \textit{i} if the player has previously engaged with the game.

\subsubsection{\textbf{Problem Statement}}
The central objective of this study is to develop a system for game recommendations for each player $u\in \mathcal{U}$ by generating a top \textit{K} list of games, denoted as $\mathcal{I}_K^{(u)}=\{i_{1}^{(u)},i_{2}^{(u)},...,i_{K}^{(u)}\}$, which player \textit{u} has not yet engaged with. Furthermore, this study introduces a multifaceted task that integrates considerations of both accuracy and diversity, requiring that the recommended lists of $K$ games not only match the interests of the players but also maintain sufficient diversity.

\subsection{Discovering the Long-tail Nature of Video Games}
\label{sec: long-tail nature}

In various real-world recommendation scenarios, such as short videos, movies, and products, user feedback on items is a critical reflection of user interest. This feedback is often quantified through metrics such as user clicks, downloads, and purchase counts, serving as proxies for item popularity within the recommendation framework. The long-tail effect describes the significant imbalance in the popularity distribution among items. Specifically, a limited number of items capture the majority of user engagement, while a substantial proportion of items, commonly referred to as ``long-tail items'', receive minimal attention.

The long-tail effect can adversely affect user satisfaction with recommender systems by impairing their ability to model user preferences effectively. This phenomenon arises because popular items, characterized by extensive historical interactions with a broad user base, intensify the high-dimensional similarity representations between these items and users. Consequently, popular items tend to dominate the ranking process, overshadowing long-tail items. Thus, a significant challenge for recommender systems is disrupting this cycle and ensuring long-tail items are recommended with sufficient frequency despite the prevailing long-tail effect.

We highlight that the evident long-tail effect in video games is a crucial motivation and significant practical foundation for this study. Specifically, the following intuitive depiction and quantitative analysis of the long-tail distribution observed on the Steam platform provide an illustration of the long-tail phenomenon.

\begin{figure}[htbp]
  \begin{minipage}{0.35\textwidth}
    \centering
    \includegraphics[width=\textwidth]{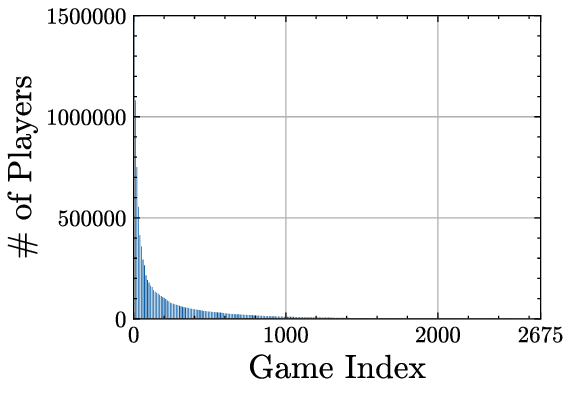}
    \captionsetup{labelformat=empty}
    \caption*{(a) Games sorted in descending order of number of players.}
  \end{minipage}%
  \hspace{20mm}
  \begin{minipage}{0.31\textwidth}
    \centering
    \vspace{0.5mm}
    \includegraphics[width=\textwidth]{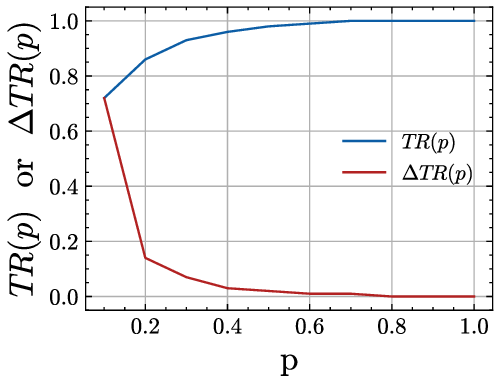}
    \captionsetup{labelformat=empty}
    \caption*{~(b) A quantitative depiction based on $TR(p),~\Delta TR(p)$.}
  \end{minipage}
\caption{The long-tail phenomenon in the number of players across all games within Steam platform, with quantitative analysis via $TR(p)$ and $\Delta TR(p)$.}
\label{fig:long-tail phenomenon}
\end{figure}

\textbf{Intuitive depiction}: As demonstrated in Fig. \ref{fig:long-tail phenomenon}(a), games are organized in descending order based on the number of players to effectively illustrate the long-tail phenomenon. This phenomenon is characterized by (1) a \textbf{limited number} of highly popular games, often referred to as the \textbf{``head''}, which dominate the market, and (2) a significantly \textbf{larger number} of less popular games, known as the \textbf{``long tail''}, each commanding smaller individual shares.

\textbf{Quantitative analysis}: Building on the concept of long-tail phenomenon, we propose the statistic Top Ratio $TR(p)$ to quantitatively capture the long-tail characteristics of video games. This statistic is defined as:
  \begin{align}
  TR(p) &= \frac{\# \text{ of top-$p$ popular games' players}}{\# \text{ of all games' players}},
\end{align}  
where $p$ is a fraction in [0.1,1] with increments of 0.1. It is evident that $TR(p)$ represents the cumulative share of the top-$p$ most popular games on Steam; thus, a larger $TR(p)$ for a smaller $p$ indicates a more pronounced long-tail effect. Additionally, we introduce $\Delta TR(p)$, defined as $TR(p)-TR(p-0.1)$, to show the growth of $TR(p)$ as $p$ increases.

The $TR(p)$ and $\Delta TR(p)$ presented in Fig. \ref{fig:long-tail phenomenon}(b) illustrate the significant long-tail property of Steam video games. (1) Even when \textbf{$\boldsymbol{p}$} is \textbf{small} (specifically, limited to 0.3), \textbf{$\boldsymbol{TR(p)}$} achieves a \textbf{high} value of 0.93, indicating that the most popular games account for a substantial majority of the total player base. (2) As $\boldsymbol{p}$ \textbf{increases}, the incremental gain in $\boldsymbol{TR(p)}$ \textbf{diminishes} sharply. These two observations quantitatively provide an intuitive understanding of the long-tail distribution characteristic of Steam video games.





\subsection{Discovering the Categorical Semantics of Video Games}

\label{Section: Discovering the Categorical Semantics of Video Games}

The categorical information of games is intricately linked to finer details of game content such as visual and musical styles, containing rich semantics and therefore profoundly influencing players. This enables category information to serve as a form of prior knowledge and empower the inference that the representations of games affiliated with similar categories should be positioned closer together in the high-dimensional space, and vice versa, which can be classified as one of the common ideas for Content-Based Filtering (CBF) \cite{cbf1, cbf2, cbf3}.

Following such ideas, GNNs have become one of the most commonly used basic frameworks in the field of recommender systems, one of the reasons being that they can explicitly exploit the above-mentioned category information to learn game representations in a spatial way, including by building edges based on the category information. Specifically, a general approach \cite{Zheng_2021,Yang_2022, cheng2017learning} is to establish a connection between any two items on the item graph based on a specific category, provided that they share similarities in this category. For example, RIP-Trilogy (casual, indie) and RoboBlitz (action, indie) are linked via the shared ``indie'' category.

However, the above strategy is limited by the uncertain reliability of category semantics\textemdash when category similarity fails to reflect true content similarity, it introduces noise, degrading recommendation quality. To mitigate this, our previous work \cite{cpgrec} introduced strict connections, requiring item pairs to share similarities in two categories to ensure stronger content alignment, in contrast to the raw connections used in conventional approaches. For instance, Counter-Strike: Source and Day of Defeat: Source (both action games developed by Valve) exhibit high content similarity, unlike Amnesia: The Dark Descent, which, despite being categorized as action, has distinct horror-themed graphics and puzzle-based gameplay.

\begin{figure}[htbp]
    \begin{minipage}{0.330\textwidth}
    \centering
    \includegraphics[width=\textwidth]{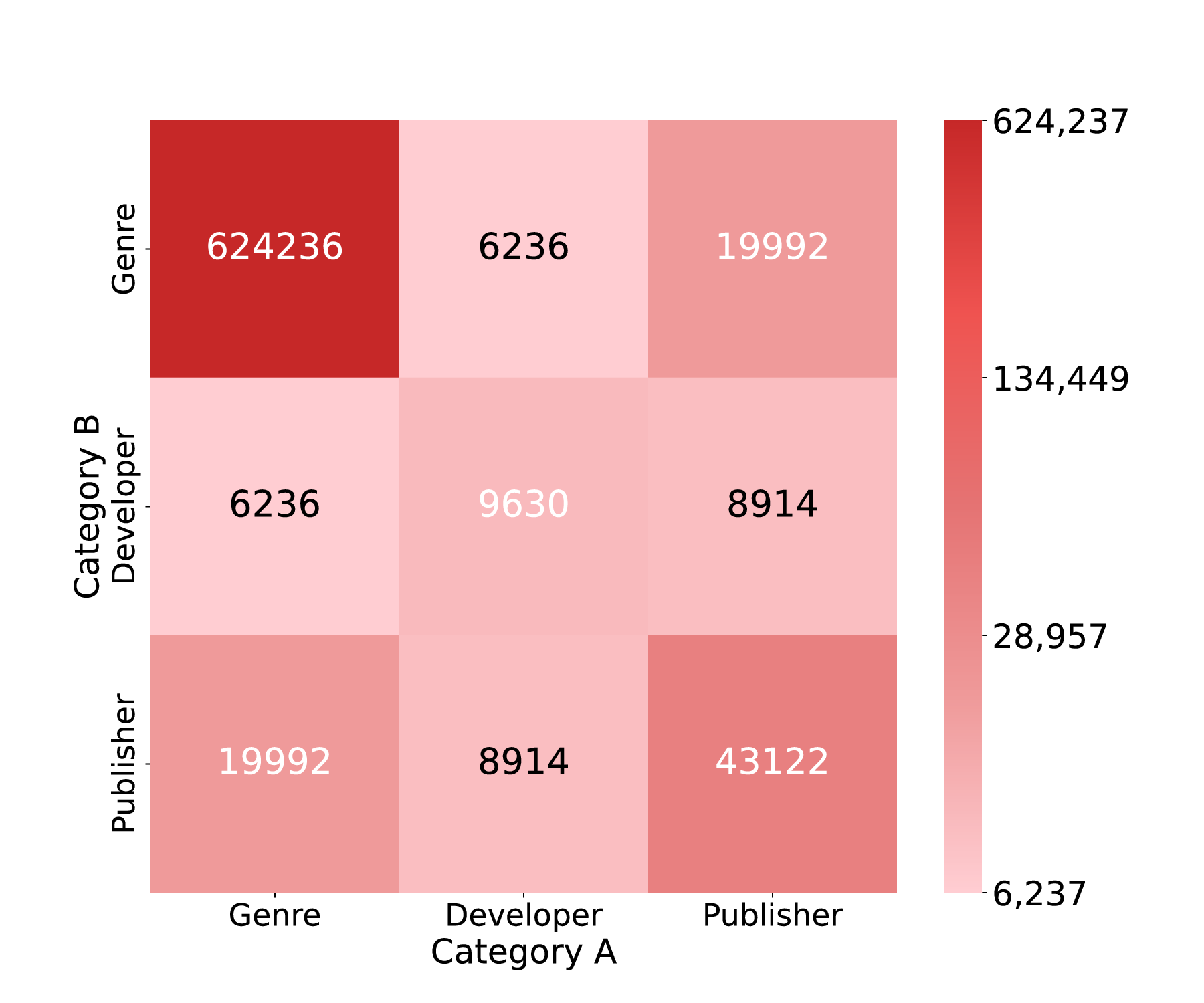} \captionsetup{labelformat=empty}
    \caption*{(a) Quantity of edges}
    \end{minipage} %
    \begin{minipage}{0.330\textwidth}
    \centering   \includegraphics[width=\textwidth]{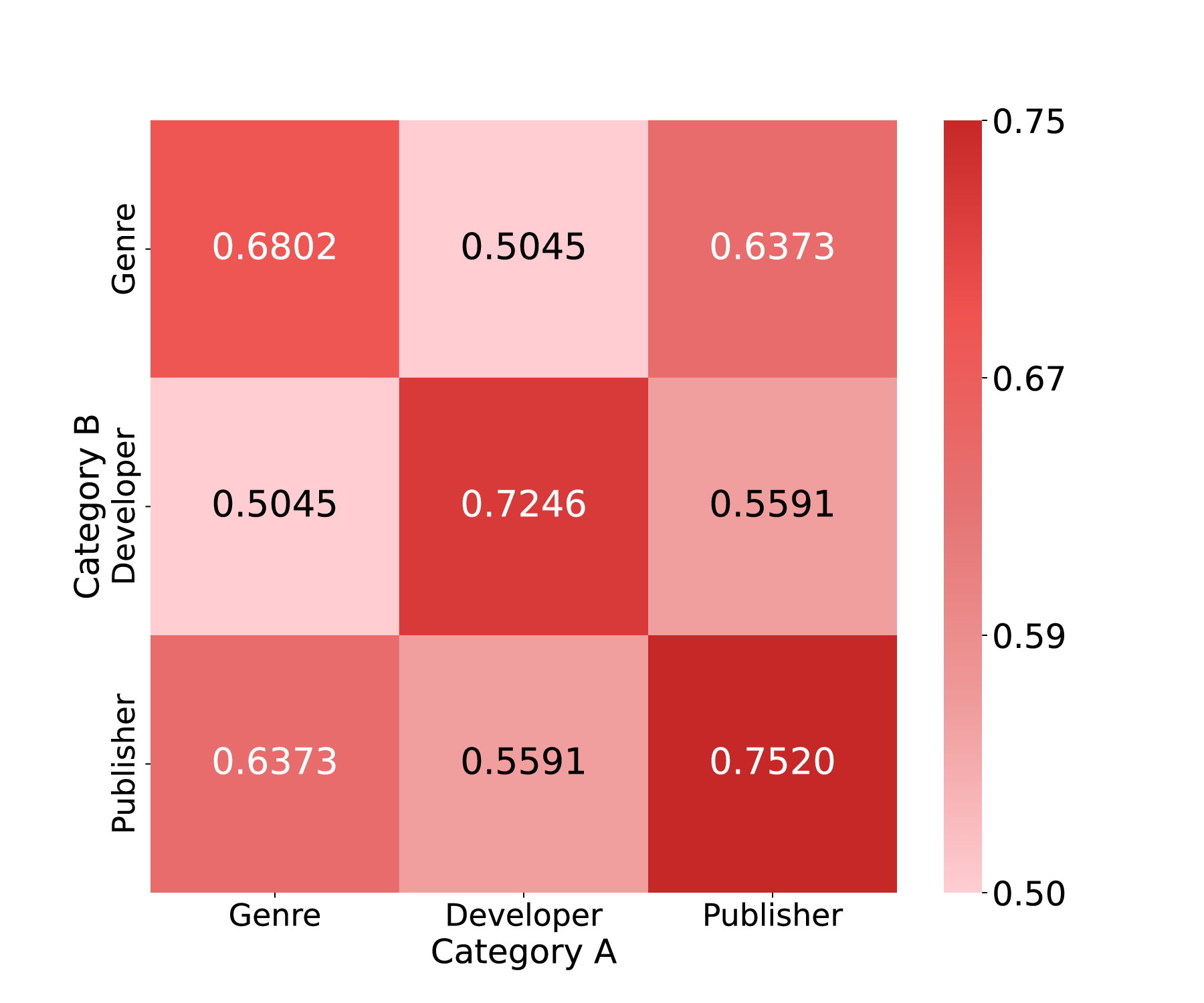}   \captionsetup{labelformat=empty}
    \caption*{(b) Average euclidean distance}
    \end{minipage} %
    \begin{minipage}{0.330\textwidth}
    \centering  \includegraphics[width=\textwidth]{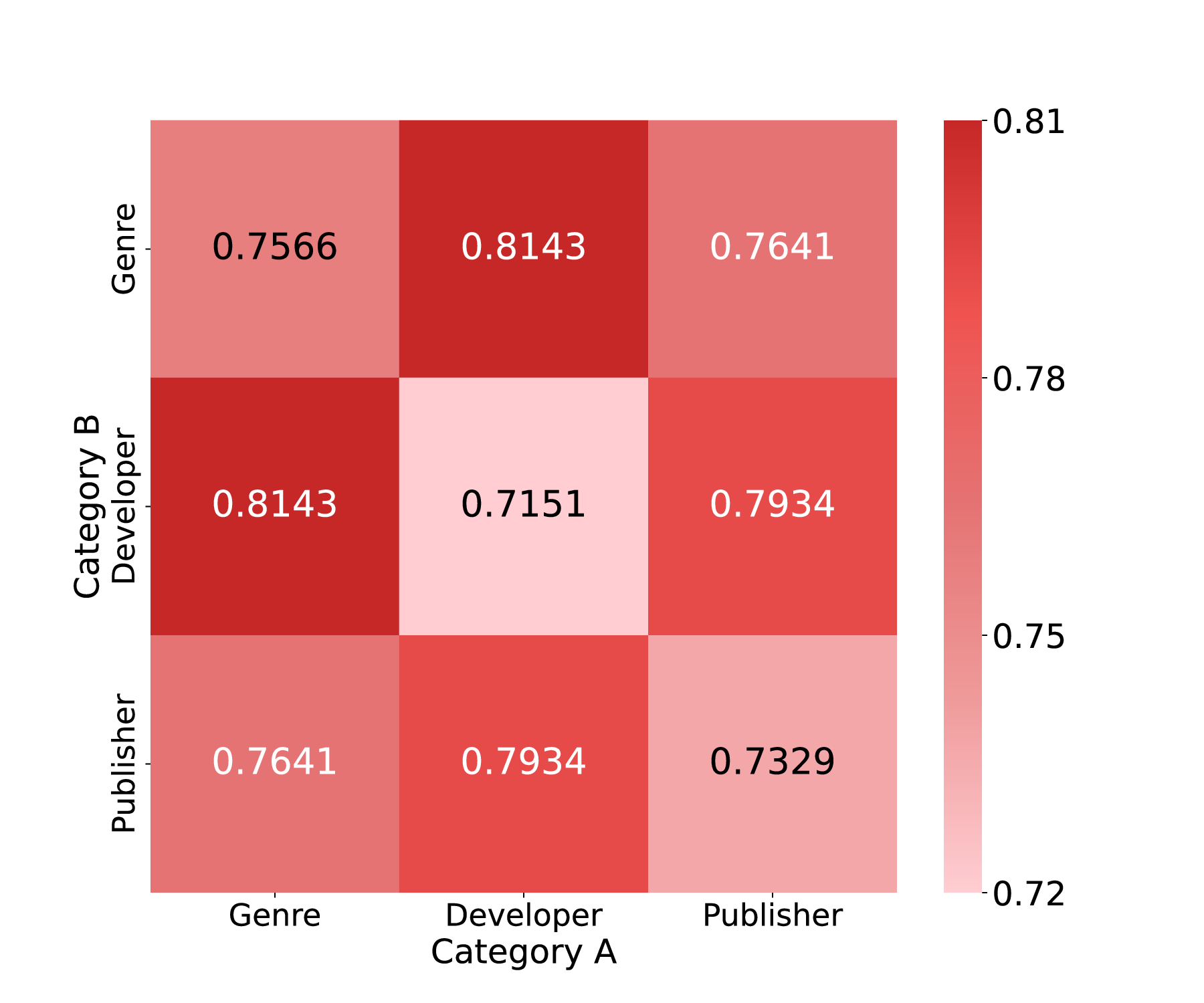} \captionsetup{labelformat=empty}
    \caption*{(c) Average cosine similarity}
    \end{minipage} %
\caption{Figure (a) presents the number of edges in the Steam game network. Figures (b) and (c) show the average Euclidean distance and cosine similarity of game description embeddings, respectively. These descriptions are generated using a large language model (LLM), and embeddings are obtained via an embedding model. The main diagonal elements reflect the Steam game network based on raw connections, while other elements are based on strict connections.}
\label{fig:data matrices}
\end{figure}

To validate the effectiveness of strict connections, we analyze the Steam network by comparing edge quantity, Euclidean distance, and cosine similarity of game description embeddings under raw and strict connections. Descriptions are generated using Qwen2.5 \cite{Qwen2.5} for contextual understanding and M3-Embedding \cite{BAAI-bge-m3} for capturing semantic meaning—both chosen for their strong performance, accessibility, and cost-effectiveness. As shown in Fig. \ref{fig:data matrices}(a), introducing a second category condition significantly reduces edge quantity, filtering out noisy connections and enhancing CBF signal quality. In Fig. \ref{fig:data matrices}(b), the larger main diagonal elements indicate that strict connections link games with more similar content, as they exhibit smaller Euclidean distances. Likewise, Fig. \ref{fig:data matrices}(c) shows larger main diagonal elements, confirming higher cosine similarity and stronger content alignment under strict connections.

\section{Review of CPGRec}

This section reviews our model CPGRec proposed in conference work \cite{cpgrec}, which is composed of four essential modules: Stringency-improved Game Connection (SGC), Connectivity-enhanced Neighbor Aggregation (CNA), Popularity-guided Edges and Nodes Reweighting (PENR), and Combined Training with Negative-sample Score Reweighting (NSR). As these methods have been detailed in prior work \cite{cpgrec}, this study provides only a concise overview.

\subsection{Stringency-improved Game Connection}
\label{Section: Stringency-improved Game Connection}
As detailed in Section \ref{Section: Discovering the Categorical Semantics of Video Games}, category information is rich in semantics. Since games within the same category are expected to share similar content, effective game representations should capture this insight. Category information thus serves as a key semantic guidance in video game modeling.

Based on this insight, the concept of strict connections is proposed to preserve connections only between games that share at least two categories among genre, developer, and publisher. This approach provides a more solid guarantee for the rationality of our inference from category similarity to the similarity of learned representation. Subsequently, LightGCN \cite{he2020lightgcn} with category-wise self-attention \cite{vaswani2017attention} is employed to learn the representation, which is defined as follows:
\begin{equation}
e_{i}^{Ca} = \text{Graphwise Attention}(e_{i}^{g\&d}, e_{i}^{g\&p}, e_{i}^{d\&p}),
\end{equation}
where $e_{i}^{g\&d}$, $e_{i}^{g\&p}$, and $e_{i}^{d\&p}$ denote the learned representations obtained using LightGCN on $\mathcal{G}^{g\&d}$, $\mathcal{G}^{g\&p}$, and $\mathcal{G}^{d\&p}$, the three game graphs with strict connections defined in Section \ref{sec: define raw graph}, respectively.

\subsection{Connectivity-enhanced Neighbor Aggregation}
\label{Section: Connectivity-enhanced Neighbor Aggregation}

To enhance diversity, we promote message propagation and aggregation across different game categories on a game graph constructed by strengthening the game graph's connectivity. Specifically, we firstly construct a game graph $\mathcal{G}^{Co}$ where neighboring games are only required to share at least one common category but are allowed to differ significantly in others. For instance, adjacent games might share the same genre but have different developers and publishers. Second, to boost cross-category message interaction, we apply multi-layer LightGCN for message propagation and aggregation within $\mathcal{G}^{Co}$, and use an attention mechanism to weight embeddings from different layers, reducing over-smoothing risks. Third, we introduce a layer-wise reweighting parameter:
\begin{equation}
    w_{l} = 1-(k-l)\beta,    
\end{equation}
where $w_{l}$ is the reweighting parameter for the $l$-th layer, $k$ is the number of LightGCN layers, and $\beta$ is the decay parameter assigning higher weights to deeper layers' embeddings. Specifically, $w_{l}$ weights the $l$-th layer's game embedding to emphasize those from deeper layers, which capture messages from more distant and diverse games. Formally, the embedding of game $i$ is obtained by:
\vspace{-1mm}
\begin{equation} 
e_{i}^{Co} = \text{Layerwise Attention}(w_{1}e_{i}^{(1)},w_{2}e_{i}^{(2)},...,w_{k}e_{i}^{(k)}), 
\end{equation}
where $e_{i}^{(l)}$ is the output embedding of game $i$ from the $l$-th layer.

\subsection{Popularity-guided Edges and Nodes Reweighting}
\label{Section: Popularity-guided Edges and Nodes Reweighting}

To emphasize diversity, the aim of this module is to create a situation where, although under normal circumstances, long-tail game nodes on player-game bipartite graphs have insufficient connectivity to influence the final learned player representation. In contrast, with the design of this link, in the alternating process of message propagation and aggregation, once messages from long-tail games flow through popular games, they can be widely disseminated to player nodes with the help of the latter's advantageous connectivity; at the same time, the influence of popular game nodes themselves should also be limited. To achieve this, we increase the weights of out edges from popular games and those of long-tail game nodes, while decreasing the weights of popular game nodes. Formally, the edge weight function $\Theta^{Po}_{e}(\cdot)$ is specified as follows:
\begin{equation}
\label{eq: ENW}
    \Theta^{Po}_{e}(i) = \begin{cases}
    \Theta_{e}^{hot} & i \in \mathcal{I}_{hot} \\
    1 & i \notin \mathcal{I}_{hot}
\end{cases},
\end{equation}
and similarly a node weight mapping $\Theta^{Po}_{n}(\cdot)$ is defined as:
\vspace{-2mm}
\begin{equation}
\label{eq: ENW2}
    \Theta^{Po}_{n}(i) = \begin{cases}
    \Theta_{n}^{hot} & i \in \mathcal{I}_{hot} \\
    1 & i \notin \mathcal{I}_{hot}\cup\mathcal{I}_{cold} \\
    \Theta_{n}^{cold} & i \in \mathcal{I}_{cold}
\end{cases},
\end{equation}
where $\mathcal{I}_{hot}$ and $\mathcal{I}_{cold}$ are respectively the sets of popular games and long-tail games. Then, multiple graph convolution layers which integrate the aforementioned reweighting mechanism are characterized by:
\begin{align}
\label{eq:reweighted convolution layer}
    {e}_{u}^{(l+1)} &= \frac{1}{\sqrt{|\mathcal{N}_{u}|}\sqrt{|\mathcal{N}_{u}|}}e_{u}^{(l)} +
    \sum_{i \in \mathcal{N}_{u}}\frac{\Theta^{Po}_{e}(i)\Theta^{Po}_{n}(i)}{\sqrt{|\mathcal{N}_{u}|}\sqrt{|\mathcal{N}_{i}|}}{e}_{i}^{(l)},
    \\
    {e}_{i}^{(l+1)} &= \frac{\Theta^{Po}_{n}(i)}{\sqrt{|\mathcal{N}_{i}|}\sqrt{|\mathcal{N}_{i}|}}e_{i}^{(l)} + 
    \sum_{u \in \mathcal{N}_{i}}\frac{1}{\sqrt{|\mathcal{N}_{i}|}\sqrt{|\mathcal{N}_{u}|}}{e}_{u}^{(l)}.
\end{align}
Here, $ \mathcal{N}_{i} $ denotes the set of neighbors for game $ i $, while $ \mathcal{N}_{u} $ denotes the set for player $ u $. The terms $ e_{i}^{(l)} $ and $ e_{u}^{(l)} $ are the embeddings output by the $ l $-th layer of LightGCN for game $ i $ and player $ u $, respectively. After processing through $ k $ layers, the embeddings for player $ u $ and game $ i $ in this module are determined by the following calculation:

\begin{equation}
    e_{u}^{Po} = e_{u}^{(k)}, 
    e_{i}^{Po} = e_{i}^{(k)}.  
\end{equation}

\subsection{Combined Training with Negative-sample Score Reweighting}
\label{Section: Combined Training with Negative-sample Score Reweighting}

The key consideration in delivering a satisfying gaming experience to players is achieving a balance between accuracy and diversity. The first module (SGC) and the last two modules (CNA, PENR) proposed above form the accuracy-oriented and diversity-oriented modules of our framework, respectively. To integrate these aspects, CPGRec adopts a weighted combination method governed by learnable parameters $w_{Ca}, w_{Co}, w_{Po}$ to obtain the final representations for each player $u$ and game $i$, which is defined as
\begin{equation} \label{eq:gamma}
\begin{aligned}
    & e_{u} = e^{Po}_{u}, \\
    & e_{i} = w_{Ca}e^{Ca}_{i} + w_{Co}e^{Co}_{i} + w_{Po}e^{Po}_{i}, \\
    & w_{Ca} + w_{Co} + w_{Po} = 1,
\end{aligned}
\end{equation}
where $w_{Ca}, w_{Co}, w_{Po}$ are the weights for the embeddings of game $i$ obtained in modules SGC, CNA, and PENR, respectively. By adjusting these weights, we can adapt the model to meet various requirements. Moreover, CPGRec further enhances both accuracy and diversity by refining the rating scores for negative samples to facilitate more efficient learning from samples, which is expressed as:
\begin{equation}  
    \overset{\sim}{L}_{BPR} = 
    -\sum_{\substack{
    u\in\mathcal{U}, 
    i\in\mathcal{I}(u), \\
    j\notin\mathcal{I}(u)
  }} \log\ \sigma(r_{u,i}-\overset{\sim}{r}_{u,j})+\lambda_{norm} \| \Theta \|_{2}^{2},
\end{equation}
where $\sigma(\cdot)$ is the sigmoid function, $r_{u,i} = e_{u} \cdot e_{i}$ is the rating score of positive sample $i$, \textcolor{black}{$\lambda_{norm}$ serving as a hyper-parameter is the weight of the normalization term (following the general setting of existing recognized works \cite{he2020lightgcn,Yang_2022, Yang_2023}, it is set to be fixed)}, and
\begin{equation}
\label{eq:negative score reweighting}
    \overset{\sim}{r}_{u,j} = m \cdot \sigma(r_{u,j}) \cdot r_{u,j},
\end{equation}
is the reweighted rating score of negative sample $j$, with $m$ controlling the reweighting intensity.

From the perspective of enhancing \textbf{accuracy}, a negative sample predicted with a high rating score could be deceptive, \textit{i.e.}, it can easily be mistaken for a positive sample. Thus, the increased loss through negative-sample score reweighting compels an improvement in the model's recognition capabilities, aiming to achieve more accurate predictions in such cases.

From the perspective of promoting \textbf{diversity}, a low score assigned to a negative sample implies substantial dissimilarity between that sample and players in terms of their representations. This could indicate limited interactions due to the game's poor exposure to players, suggesting it might be a long-tail game. By increasing the rating score, this module enhances their chances of being recommended.

\section{Motivation for New Module}

\subsection{Preliminary: Smoothness Nature of GCNs}
In the field of graph learning, the concept of smoothness refers to a property of GNN models (\textit{e.g.}, Graph Convolutional Network (GCN)). It means that the representations learned by a GNN-based model with smoothness of nodes that are close to each other on a graph (\textit{i.e.}, connected or in close proximity) tend to be similar (\textit{e.g.}, their cosine similarity is much higher, or their Euclidean distance is much lower). The vast majority of GNN-based models, especially the most widely used GCN, possess the property of smoothness, which has a dual nature: \textbf{(1)} Smoothness can be leveraged to capture commonalities between nodes. When nodes are connected or part of the same neighborhood in a graph, they often share similar characteristics or roles. By enforcing smoothness, GNNs can effectively propagate and aggregate information between neighboring nodes, leading to more coherent and consistent representations; \textbf{(2)} However, excessive smoothness can lead to the problem of over-smoothness, where the representations of nodes may become too similar, losing the distinctiveness needed for effective learning and prediction.

In CPGRec, smoothness is utilized to capture the commonalities between games sharing similar categories in the module SGC, based on the belief that games with similar categories are expected to have similar content, effectively enhancing the recommendation accuracy. Nonetheless, as mentioned earlier, the direct application of GCN layers may lead to the over-smoothness issue. To better leverage smoothness while mitigating over-smoothness, it is essential to understand the smoothness property of GCN.

From a spatial domain perspective, smoothness is relatively straightforward to understand, as feature interactions occur through message propagation and aggregation between adjacent nodes, resulting in greater similarity among neighboring nodes. However, this intuitive perspective is insufficient, as it fails to adequately explain why certain GNNs, particularly those designed from the spectral domain, hold the potential to address over-smoothness. Therefore, a spectral perspective is provided in Appendix \ref{preliminary: smoothness} for a deeper understanding of smoothness.


\subsection{Motivation 1: Smoothness of GCN Layer within CPGRec}
\label{impair}

The GCN \cite{kipf2017semisupervised, he2020lightgcn} serves as a foundational framework in CPGRec, facilitating representation learning for players and video games within game graphs and the bipartite graph. Nevertheless, it is essential to recognize that the role of GCN in CPGRec is not uniformly advantageous.

On one hand, the smoothness property of GCNs enhances game representation learning by effectively capturing the rich semantics of various categories in the context of the SGC module. Discussions in Section \ref{Section: Discovering the Categorical Semantics of Video Games} introduce the rationality of the belief that categorically similar games should share similar representations, which serves as a principle that offers guidance to game modeling, and can be effectively implemented through the smoothness property of GCNs. 

Conversely, the smoothness property introduces the risk of compromising accuracy in module PENR, which is based on the player-game graph. The tendency for representations of players and games to converge towards similarity poses a significant challenge to the effectiveness of CPGRec, as it neglects the consideration of diverse preferences and behavioral patterns among different players. Furthermore, while popularity-guided weights are introduced to enhance long-tail influence and diversify recommendations, the experimental results of the ablation study in our conference work \cite{cpgrec} have demonstrated that they may even impair accuracy, albeit to an acceptable degree. Potential reasons are that these popularity-guided weights are not conducive to learning complex interaction patterns in the player-game context, as the popularity information solely reveals nothing about the player-game interactions.





\begin{figure}[htbp]
  \begin{minipage}{0.35\textwidth}
    \centering
    \includegraphics[width=\textwidth]{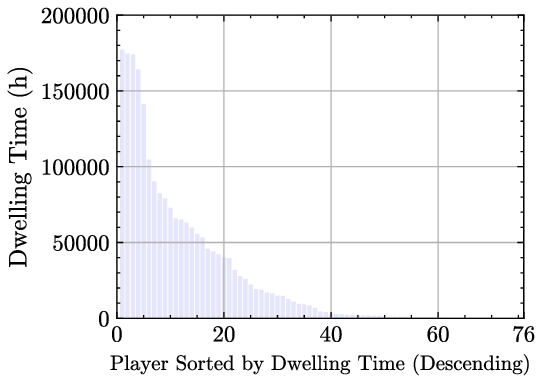}
    \captionsetup{labelformat=empty}
    \caption*{(a) Dwelling time of all players for game \textit{Electronic Super Joy}}
  \end{minipage}%
  \hspace{1cm}
  \begin{minipage}{0.35\textwidth}
    \centering
    \includegraphics[width=\textwidth]{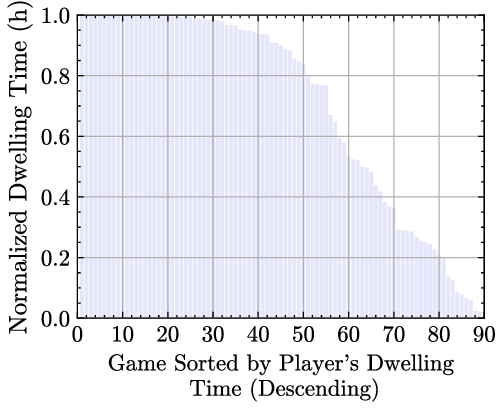}
    \vspace{-0.85cm}
    \captionsetup{labelformat=empty}
    \caption*{(b) Dwelling time of the  player \textit{320}'s for historical games}
  \end{minipage}
\caption{Two concrete examples for illustrating the disparities of historical interactions.}
\label{fig:inequality}
\end{figure}

\subsection{Motivation 2: Disparities of Historical Interactions in Player-Game Graph}
\label{section: disparity}

In this section, we conduct several data explorations to illustrate the problems arising from edge weights derived from the PENR module (introduced in Section \ref{Section: Popularity-guided Edges and Nodes Reweighting}) in the context of the player-game bipartite graph. This analysis reveals the underlying causes of over-smoothness and highlights the motivations and foundations of this study. Specifically, we discuss the unequal importance and rich semantics of historical interactions from two perspectives:
\begin{itemize}
    \item From the viewpoint of a specific \textbf{game}, the duration of engagement by historical players exhibits considerable variation, indicative of differing levels of enthusiasm among players for that game.
    \item From the viewpoint of a specific \textbf{player}, engagement time across historical games varies significantly, reflecting the player's unique preferences for different games.
\end{itemize}
To provide a more intuitive illustration, we offer concrete examples for both scenarios: from the game perspective, we use the game \textit{Electronic Super Joy} as an example; from the player perspective, we focus on the player with \textit{ID=320}. These examples are depicted in Fig. \ref{fig:inequality}.

Fig. \ref{fig:inequality}.(a) shows the dwelling time of all 76 players for the game \textit{Electronic Super Joy}. Despite their observed historical interactions with the game, dwelling times exhibit considerable variability. Notably, one player has amassed an impressive 177,260 hours of playtime, while another player has not engaged with the game at all since its acquisition. We contend that only the former's substantial engagement accurately reflects genuine interest, whereas the latter's inactivity does not.

Fig. \ref{fig:inequality}.(b) illustrates the Z-score-normalized dwelling times for the player with \textit{ID=320} across their 90 historical games. Although the player has interacted with these games, dwelling times vary significantly. For instance, after Z-score normalization, the player's engagement with certain games approaches nearly 1.0, indicating a strong alignment with their interests. In contrast, a dwelling time close to 0.0 for other games suggests minimal engagement post-purchase, leading us to classify these games as unrepresentative of the player’s preferences.

\textbf{Summary.} Considering the smoothness property of GCNs, neglecting the critical disparities in historical interactions can cause the representations of players and games to become overly similar, potentially hindering CPGRec's capability to provide more accurate recommendations. To address this issue, we propose integrating a new module called \textit{Preference-informed Edge Reweighting (PER)}, which aims to model the personal preference of each player by factoring in interaction-wise semantics, where both significant interest and disinterest are included. \textcolor{black}{Moreover, current recommender systems fail to fully utilize LLMs' potential in deeply mining players' personal preferences. To bridge this gap, we introduce the \textit{Preference-informed Representation Generation (PRG)} module,} \textcolor{black}{which leverages LLMs to enhance the modeling of players' preferences by refining game descriptions} based on average ratings (global interest) and generating player descriptions through a comparative analysis of dwelling time (personal interest) and global interest.

\begin{figure*}[t]
\vspace{-5mm}
\begin{minipage}{\textwidth}
\hspace{-7mm}
\includegraphics[width=1.1\textwidth]{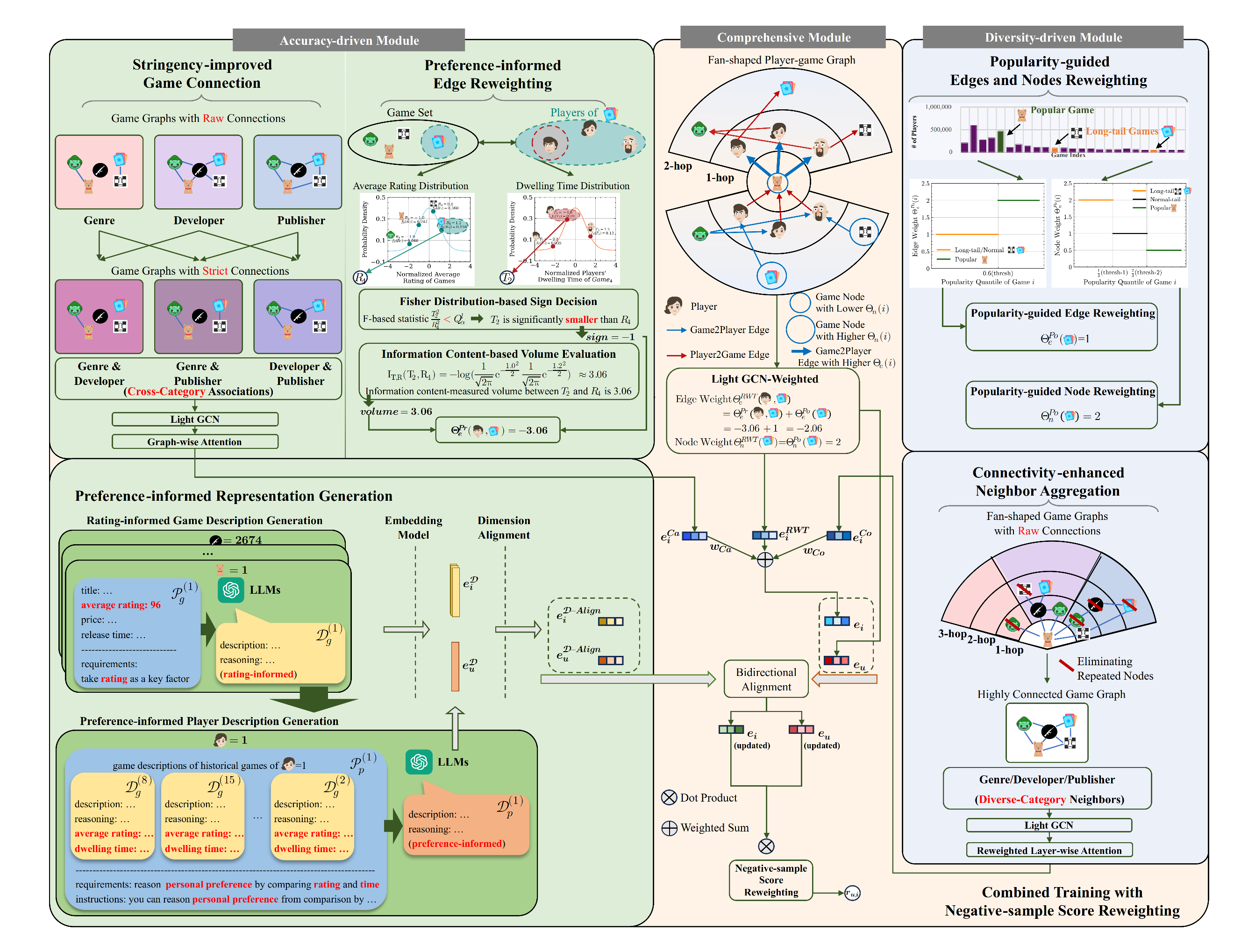}
\vspace{-5mm}
\caption{Illustration of the CPGRec+ framework, comprising three modules: Accuracy-driven module (Stringency-improved Game Connection, Preference-informed Edge Reweighting, and Preference-informed Representation Generation), Diversity-driven module (Connectivity-enhanced Neighbor Aggregation, Popularity-guided Edges and Nodes Reweighting), and Comprehensive module.}
\label{fig:framework_tois}
\end{minipage} \vspace{-5mm}
\end{figure*}

\section{CPGRec+: Towards \textcolor{black}{Preference-informed} Recommendation}

\textcolor{black}{In this section, we begin by analyzing the interest and disinterest in our study.} 
Subsequently, we propose two novel modules, PER and PRG, to capture these preferences with edge reweighting and LLMs respectively. \textcolor{black}{Both modules address the challenge of interaction disparities, while PER further mitigates the risk of over-smoothing inherent in CPGRec.} Finally, we show the overview of the improved CPGRec+ model and its detailed analysis.

\textbf{Player's preference.} A player's personal preference is manifested in the discrepancies between his or her preference and those of the broader player base. Specifically, each historical interaction potentially reflects a player's personal preference from two perspectives: \textbf{(1) interest}: a player may exhibit greater interest than the general player base on the observed interactions with a non-popular game, reflecting his or her personal \textbf{interest} in that game, in stark contrast to the general historical players of this involved game; \textbf{(2) disinterest}: conversely, a player may demonstrate significantly less interest than the general player in a popular game, indicating his or her disinterest. To this end, PER and PRG are proposed in this paper to detect these informative interactions, and leverage them to refine the representations of games and players.

\subsection{New Module: Preference-informed Edge Reweighting}
\label{new module: per}
A player-game bipartite graph that neglects the disparities of observed player-game interactions is an under-improved modeling of player-game history relationships. This is because it does not fully consider real-world situations, leading to suboptimal collaborative filtering signals. Furthermore, such neglect may trigger problems arising from the inherent smoothness of GCN, which results in increased similarity among node representations. These factors ultimately lead to suboptimal modeling for players and games. \textcolor{black}{To address this problem, we propose the PER module, which solves this problem in two steps. First, inspired by existing works in anomaly detection field to address over-smoothing problem \cite{h2fd, ghrn, caregnn}, along with recent advancements for identifying noisy signals within complex data~\cite{pyliu2, pyliu3}, we design a sign decision mapping for player-game edges within the bipartite graph to differentiate between interest and disinterest of player. Then, we further introduce information content \cite{info_content} to carefully quantify the significant preferences identified in the previous step.}

\subsubsection{Fisher Distribution-based Sign Decision}
We propose introducing a sign decision mapping for edges within the player-game bipartite graph, which differentiates between interest and disinterest as revealed by each historical interaction. Such sign mapping for messages can be seamlessly integrated into graph convolutional layers and has demonstrated efficacy in mitigating over-smoothness in the field of data mining, such as anomaly detection \cite{h2fd, ghrn, caregnn}.

Specifically, we detect those interactions that show significant player preferences (including both interest and disinterest) based on the involved dwelling time and the average ratings of the involved games. Then, we assign a positive sign to edges representing interest and a negative sign to those indicating disinterest. This approach allows us to convey messages with a positive sign through the former and messages with a negative sign through the latter. As a result, we expect a reduction in the high-dimensional representation similarities between players and the games they show significant disinterest in, while simultaneously increasing similarities with games they show significant interest in.

To achieve this, we construct the sign mapping by employing both the dwelling time and average ratings as metrics. The reason for adopting them is that dwelling time revealed by each single historical interaction reflects this particular player's interest in this specific game; in contrast, the average rating of this game reflects the interest of general players in this game. By comparing these two indicators, they jointly reflect the personal preference of this player indicated by each historical interaction. However, this raises two critical questions:


\textcolor{black}{
\begin{itemize}
    \item The dwelling time and average ratings should be from different spaces, prohibiting comparisons between them (our goal). How to transform dwelling time and ratings into the same space?
    \item How can we define a method that compares dwelling time and average ratings to effectively infer the personal preferences of each player?
\end{itemize}
}

\textcolor{black}{
To solve the \textbf{first} question, we propose to introduce the Box-Cox transformation \cite{boxcox1}. Box-Cox transformation is a statistical method that can effectively map any input data into an approximately normal distribution while maintaining its relative relationship, and has been widely employed in various fields, such as anomaly detection \cite{boxcox_a4, boxcox_a3}, fuzzy systems \cite{boxcox_a2}, and Monte Carlo denoising \cite{boxcox_a1}. Then, we nest a Z-score normalization to further turn the data output by Box-Cox transformation into approximately standard normal distribution, which eventually maps both the dwelling time and average ratings to the same space (standard normal space).
\\
Formally, for any input data  $y$ (\textit{i.e.}, dwelling time or average ratings), the transformed data $y^{(\lambda_{BC})}$ is computed by:
\begin{align}
    y^{(\lambda_{BC})}=\begin{cases}
        \frac{y^{\lambda_{BC}}-1}{\lambda_{BC}} & \text{if}~~ \lambda_{BC} \neq 0, \\
        \ln(y) & \text{if}~~ \lambda_{BC} = 0,\\
    \end{cases}
\end{align}
where $\lambda_{BC}$ is a hyper-parameter that controls the skewness and kurtosis of the transformed distribution. \textcolor{black}{It is worth clarifying that $\lambda_{BC}$ is a hyperparameter that remains fixed during supervised learning rather than a learnable parameter, determining how close the transformed data are to a normal distribution. A range of methods have been proposed to estimate the optimal value of $\lambda_{BC}$ \cite{boxcox1}, of which we leverage the implementation of Scipy \footnote{https://docs.scipy.org/doc/scipy/reference/stats.html} to obtain it automatically using maximum likelihood.} Subsequently, the Z-score transformation is introduced to further turn $y^{(\lambda_{BC})}$ into $z$ by: 
\begin{align}\label{z=}
    z = \frac{y^{(\lambda_{BC})}-\mathbb{E}[y^{(\lambda_{BC})}]}{\sqrt{\mathrm{Var}[y^{(\lambda_{BC})}]}},
\end{align}
which approximately follows the standard normal distribution. $\mathbb{E}$ and $\mathrm{Var}$ are the expectation and variance of data $y^{(\lambda_{BC})}$, respectively. Moreover, we demonstrate the effectiveness of our proposed composite mapping (Box-Cox transformation followed by Z-score normalization) in transforming dwelling time and average ratings into the same standard normal space by employing One Sample Kolmogorov-Smirnov Test \cite{KS-test} in Section \ref{Section: Validation of the Statistical Basis of PER Module}.
}

The standard normality of the transformed data inspires us to design a comparison method to solve the \textbf{second} problem: modeling the personal preferences of each player. Specifically, Fisher \cite{fisher} proposed the concept of Fisher distribution, which can be regarded as the square of the ratio of two independent standard normal distributions, and the distribution characteristics of Fisher distribution are clearly known. This allows us to use the square of the ratio between the transformed dwelling time and the average rating as the result of the comparison, and use Fisher distribution as the basis for measuring the significance of this comparison result, to obtain a qualitative comparison result.

Formally, given that the observed dwelling time of player $u$ on game $i$ is denoted as $t_{u,i}$, and the average rating of game $i$ is denoted as $r_i$, we know that the mapped dwelling time $T$ for each video game and the mapped average ratings $R$ of all games respectively follow a standard normal distribution, which are expressed as:
\begin{align}
 f_{T}(t_{u,i}) &= \frac{1}{\sqrt{2\pi}}\exp(-t_{u,i}^{2}),~t_{u,i}\in \mathcal{R},\\
 f_{R}(r_{i}) &= \frac{1}{\sqrt{2\pi}}\exp(-r_{i}^{2}),~r_{i}\in \mathcal{R}.
\end{align}
According to the definition of Fisher distribution, the statistic \( F = \frac{T^{2}}{R^{2}} \) follows a Fisher distribution with parameters $(d_1,d_2)=(1,1)$ \cite{fisher}, whose cumulative distribution function of $F$ is defined as:
\begin{align}
 P(F\leq x)&=\mathcal{I}_{\frac{x}{x+1}}(\frac{1}{2},\frac{1}{2}),
\end{align}
where $\mathcal{I}$ is the regularized incomplete beta function as:
\begin{align}
 \mathcal{I}_{x}(a,b)&=\frac{B(x;a,b)}{B(a,b)},\\
 B(x;a,b) &= \int_{0}^{x}t^{a-1}(1-t)^{b-1}dt.
\end{align}
Thus, we can directly obtain the upper quantile $Q_{\alpha} \in \mathcal{R}$ of $F$ (hyper-parameter $\alpha$ controls the significance level) and take it as a clear criterion for whether $T^2$ is significantly greater than $R^2$, considering
\begin{align}
F \gg Q_{\alpha} \Longleftrightarrow T^2 \gg R^2, F=\frac{T^2}{R^2}.
\end{align}
It is noteworthy that the condition $F \gg Q_{\alpha}$ highlights the two very different cases $T\gg0, R\approx 0$ and $T\ll0, R\approx 0$, which respectively show the significant \textbf{interest} and \textbf{disinterest} of the player revealed in the particular interaction: 
\begin{itemize}
    \item $T\gg0, R\approx 0$ respectively indicate that the player has a strong interest for this game (indicated by a fairly long dwelling time) while the global interest of all players for this game is at a mediocre level (indicated by the average rating of the game), which together reflect the player's significant \textbf{interest} in the game.
    \item $T\ll0, R\approx 0$ respectively indicate that the player has a strong disinterest for this game (indicated by a fairly short dwelling time) while the global interest of all players for this game is at a mediocre level (indicated by the average rating of the game), which together reflect the player's significant \textbf{disinterest} in the game.
\end{itemize}
This inspires us to construct the preference-informed sign mapping for the historical interaction between player $u$ and game $i$ by:
\begin{align}
    sign^{Pr}_{e}(u,i) &= \begin{cases}
        1 & \text{if} ~~F=\frac{t_{u,i}^{2}}{r_{i}^{2}} > Q_{\alpha}~~\text{and}~~t_{u,i}>0,\\
        -1 & \text{if} ~~F=\frac{t_{u,i}^{2}}{r_{i}^{2}}> Q_{\alpha}~~\text{and}~~t_{u,i}\leq 0,\\
        0 & \text{else},\\
    \end{cases}
\end{align}
which first determines whether each historical interaction shows a significant personal preference, and if so, determines whether the preference is interest or disinterest: if it is interest, a positive sign is given to increase the similarity between the player and game representations; if it is disinterest, a negative sign is given to explicitly reduce the similarity between the two.

\subsubsection{Information Content-based Volume Evaluation}

Although preference-informed sign mapping provides a macroscopic perspective on players' personal preferences, there remains a need for a more detailed quantification of the disparities present in different historical interactions.

Inspired by existing research \cite{info1, info2, info3}, we propose to introduce information content \cite{info_content} as the metric to solve this problem. As an important concept of information theory, information content $\textbf{I}(\cdot)$ quantifies the amount of surprise or unexpectedness associated with an observed event $\textbf{A}$, which is formally computed by:
\begin{align}
    \textbf{I}(\textbf{A})=-\log(P(\textbf{A})),
\end{align}
where $P(\textbf{A})$ is the probability that $\textbf{A}$ occurs. As seen, a rarer event $\textbf{A}$ has higher information content $\textbf{I}(\textbf{A})$ since its probability $P(\textbf{A})$ is low, and vice versa. This enables us to quantitatively describe the intensity of each observed significant personal preference in a similar way. Specifically, for a historical interaction involving a player with an observed dwelling time of \( t_{u,i} \) and an observed game rating of \( r_i \), the information content is defined as:
\begin{align}
    \textbf{I}^{Pr}_{e}(t_{u,i}, r_{i}) &= -\log P(T = t_{u,i}, R = r_i).
\end{align}
Given \( T \) and \( R \) are independent, we have
\begin{align}
P(T = t_{u,i}, R = r_i) = P(T = t_{u,i}) P(R = r_i)
\end{align}
and
\begin{align}
\textbf{I}^{Pr}_{e}(t_{u,i}, r_i) = -\log P(T = t_{u,i}) P(R = r_i).
\end{align}
Eventually, the preference-informed edge weight $\Theta_{e}^{Pr}(u,i)$ is defined as:
\begin{align}
    \Theta_{e}^{Pr}(u,i) = sign^{Pr}_e(u,i) \cdot \textbf{I}^{Pr}_{e}(t_{u,i},r_i),
\end{align}
which provides a detailed, interaction-focused characterization of each historical interaction. The PER algorithm is summarized in Algorithm \ref{agrm 1:per}.

\begin{algorithm}[htbp]
    \caption{Preference-informed Edge Reweighting}\label{agrm 1:per}
    \KwIn {edge set $\mathcal{E}$, game average rating mapping $R$, player dwelling time mapping $T$, significance level $\alpha$}
    \KwOut {preference-informed edge weight mapping $\Theta_e^{Pr}(\text{\textbf{e}})$ for input edge $\text{\textbf{e}} \in \mathcal{E}$}
    $Q_{\alpha}^{u} \gets \frac{1-\alpha}{\alpha}$\\
    \For{$e$ in $\mathcal{E}$}{
        $(u,i) \gets e$\\
        $R_{i},T_{u} \gets R(i),T(e)$\\
        $F \gets \frac{T_{u}^{2}}{R_{i}^{2}}$\\
        $I_{T,R}(T_{u},R_{i}) \gets -\log(\frac{1}{2\pi}\exp(-(T_{u}^{2}+R_{i}^{2}))$\\
        \If{$F > Q_{\alpha}$}{
            \If{$T_u>0$}{$\Theta_e^{Pr}(e) \gets I_{T,R}(T_{u},R_{i})$}
            \Else{$\Theta_e^{Pr}(e) \gets -I_{T,R}(T_{u},R_{i})$}
        }
        \Else{
            $\Theta_e^{Pr}(e) \gets 0$\\
        }
    }
    \Return $\Theta_e^{Pr}(\text{\textbf{e}})$\\
\end{algorithm}

\subsection{New Module: Preference-informed Representation Generation} \label{sec: new module: PRG}

\textcolor{black}{
Large language models (LLMs) have been widely used in the field of recommender systems recently due to their rich knowledge and powerful reasoning ability. However, their potential for generating informative descriptions and comprehending historical interactions has yet to be fully leveraged in uncovering players' personal preferences. To address this gap, this study first employs LLMs to generate game descriptions that emphasize players' global interests, which are indicated by the average ratings of involved games. Subsequently, LLMs are further utilized to extract player descriptions that capture individual preferences. Finally, both types of descriptions are embedded into representations to refine player and game embeddings, enhancing the overall recommendation quality.
}

\subsubsection{Rating-informed Game Description Generation}

PRG enables LLM to deeply understand the content of each game at a semantic level by constructing textual prompts that include key game information, such as the title, rating, price, and release time (since the comparison between ratings and time is not directly addressed here, the original 100-point rating scale is retained). Specifically, PRG directs the LLM to focus on deriving insights from the average rating, which reflects the preferences of the majority of players, serving as an indicator of \textbf{global interest}. This approach leverages the LLM's capabilities in reasoning, natural language understanding, and generation to create high-quality textual descriptions for each game, which later facilitates further reasoning involving the joint evaluation of rating and dwelling time.

Formally, PRG designs an input prompt $\mathcal{P}_g^{(i)}$ for each game $i\in \mathcal{I}$ by combining two key components: (1) critical textual information about the game and (2) explicit instructions. The textual information includes the game title $\alpha_{g}$, the average rating $r_g$ (as a reflection of global interest), and supplementary data such as price $p_g$ and release date $t_{g}$. The explicit requirements instruct the LLM to generate the description with particular emphasis on $r_g$, as it reflects the global interest of general player base for this game. This can be formally expressed as:

\begin{align}
    \mathcal{P}_g^{(i)}=\begin{cases}
        f_g([\alpha_g,r_g,p_g,t_g]),  ~~~\text{if $r_g,p_g,t_g$ exist},\\
        f_g([\alpha_g,\bar{r}_g,p_g,t_g]),  ~~~\text{else if $p_g,t_g$ exist},\\
        f_g([\alpha_g,r_g]),  ~~~\text{else if $r_g$ exist},\\
        f_g([\alpha_g,\bar{r}_g]),  ~~~\text{otherwise},\\       
    \end{cases}
\end{align}
where $f_g(\cdot )$ is a game-specific function that combines the textual information with the explicit requirements into a single string. Missing game ratings are substituted with the average rating of available games, while missing supplementary information is omitted as it is non-essential.

Based on this, a state-of-the-art LLM can be leveraged to generate high-quality description as textual representation in natural language for each game $i\in \mathcal{I}$, which is expressed as:
\begin{equation}   \mathcal{D}_{g}^{(i)}=\mathcal{LLM}(\mathcal{P}_g^{(i)}),
\end{equation}
where $\mathcal{D}_{g}^{(i)}$ indicates the textual description of game $i$, which highlights the global interest of players. In this study, the LLM Qwen2.5 \cite{Qwen2.5} is utilized considering its excellent ability to understand context and generate human-like text.

Additionally, building on the success of LLMs in generating user and item profiles in recommender systems \cite{llmrec1, llmrec2, llmrec3}, along with recent advancements in encoding textual semantics into numerical embeddings~\cite{pyliu1}, we integrate an embedding model $\mathcal{EMB}$ to map $\mathcal{D}_g^{(i)}$ into a numerical semantic space. This complements the game representation, as expressed by the following equations:
\begin{align}\label{eq: MLP-embed-game}
    e_i^{\mathcal{D}} &= \mathcal{EMB}(\mathcal{D}_{g}^{(i)}), \nonumber \\
    e_i^{\mathcal{D}-Align}  &=MLP^{\mathcal{D}-Align}(e_i^{\mathcal{D}}), \nonumber \\
    e_i &= MLP^{\mathcal{D}-Integ}_g([e_i,e_i^{\mathcal{D}-Align}]),
\end{align}
where $\mathcal{EMB}$ \textcolor{black}{(M3-Embedding \cite{BAAI-bge-m3} is adopted due to its effectiveness in capturing semantic meaning)} embeds the textual description $\mathcal{D}_{g}^{(i)}$ for game $i$ into the numerical semantic space to obtain $e_i^{\mathcal{D}}$, $MLP^{\mathcal{D}-Align}$ is a Multilayer Perceptron (MLP) \cite{MLP} that aligns $e_i^{\mathcal{D}}$ from the semantic space with $e_i$ from the original game representation space, and $MLP^{\mathcal{D}-Integ}_g$ updates the game representation $e_i$ by integrating $e_i^{\mathcal{D}-Align}$, encapsulating the LLM-derived knowledge about the game \textcolor{black}{(with the MLP depth set to 2 by default in this study)}. \textcolor{black}{It is acknowledged that MLP is not the only representation fusion method, as linear, gated fusion strategies are often considered as well. Therefore, to provide a more complete analysis, we provide a comparison across them in Section \ref{sec: fusion analysis}.} Furthermore, we also provide an example in Appendix \ref{appendix: prompt template} (Fig. \ref{fig: prompt template for generating game description}) that specifically shows how to generate a description for a game following the above steps.

\subsubsection{Preference-informed Player Description Generation}

Furthermore, PRG leverages the meticulously generated game descriptions $\{\mathcal{D}_{g}^{(i)}\}_{i\in\mathcal{I}}$ to construct personalized prompts aimed at deeply exploring players' preferences by reasoning from the comparison between the dwelling time and the average ratings of games. Specifically, PRG designs a personal prompt for each player, which includes the descriptions of its historical games, the personal interest of this player in these games (represented by the normalized dwelling time $t_{u,i}$) and the global interest (represented by the normalized average rating $r_i$). By explicitly instructing the LLM to \textbf{compare personal and global interests} as the primary focus, PRG compels the LLM to extract and highlight the most relevant information that reflects the player's unique preferences from the previously generated high-quality game descriptions.

It is worth noting that, since the rating and dwelling time obtained in Equation \ref{z=} have been carefully mapped to follow a standard normal distribution, they can be directly compared and serve as quantitative indicators of personal preference. This enables the LLM to interpret and understand these features more effectively. 

Formally, the structure of the input player prompt $P_p$ for player-description generation is outlined as follows:
\begin{align}
    \mathcal{P}_p^{(u)}=
        f_p([f_{hg}(\mathcal{D}_{g}^{(i)},t_{u,i},r_i)]), ~~i\in \mathcal{I}(u),
\end{align}
where $f_{hg}$ combines the generated game description $\mathcal{D}_g^{(i)}$ with both the normalized dwelling time $t_{u,i}$ and average rating $r_i$; $f_p$ integrates all the historical games of player $u$ and explicitly requires the language model LLM to infer the player's personal preferences by comparing time and average rating based on our analysis detailed in Section \ref{section: disparity} \textcolor{black}{and the transformed data obtained in Section \ref{new module: per}}. Then, with the help of LLM, we obtain the description for each player $u$ by:
\begin{equation}
    \mathcal{D}_p^{(u)} = \mathcal{LLM}(\mathcal{P}_p^{(u)}),
\end{equation}
where $\mathcal{D}_p^{(u)}$ indicates the textual description of player $u$, emphasizing the personal interest of the player. In this context, we also utilize the LLM Qwen2.5 \cite{Qwen2.5} to generate these descriptions.

Similar to Equation \ref{eq: MLP-embed-game}, we use a state-of-the-art embedding model $\mathcal{EMB}$ \textcolor{black}{(\textit{i.e.,} M3-Embedding \cite{BAAI-bge-m3})} to integrate the preference-informed player description $\mathcal{D}_p^{(u)}$ into the player representation by:
\begin{align}\label{eq: MLP-embed-player}
    e_u^{\mathcal{D}} &= \mathcal{EMB}(\mathcal{D}_{p}^{(u)}), \nonumber \\
    e_u^{\mathcal{D}-Align}  &=MLP^{\mathcal{D}-Align}(e_u^{\mathcal{D}}), \nonumber \\
    e_u &= MLP^{\mathcal{D}-Integ}_p([e_u,e_u^{\mathcal{D}-Align}]).
\end{align}
Appendix \ref{appendix: prompt template} (Fig. \ref{fig: prompt template for generating player description}) provides a specific example illustrating how to generate a description for a player following the above steps.

\subsection{Overview of CPGRec+}
In summary, the two newly introduced modules (PER and PRG) attempt to bridge the gap resulting from our proposed model CPGRec's insufficient consideration of the disparities of observed player-game historical interactions, where disparities refer to the different importance of each player's interactions in reflecting this player's personal interest. To solve this problem, both PER and PRG attempt to better model the personal preference of each player by deeply capturing the player's interest or disinterest exhibited in each historical interaction. Methodologically, PER refines the modeling of player-game interactions by reweighting edges based on the comparison between personal interest and global interests, whereas PRG leverages LLMs to generate more informative game and player representations by explicitly guiding the model to reason from both the personal and global interests as key factors in preference inference.
 
Although both the PER and PENR modules reweight the player-game edge weights, the newly proposed PER serves as an accuracy-driven module, whereas the PENR functions as a diversity-driven module. Collectively, these two modules facilitate balance-oriented recommendations through the formulation of a \textbf{balance-oriented edge weight} defined as:
\begin{equation}\label{eq:rwt combination}
\Theta_{e}^{RWT}(e) = \Theta_{e}^{Pr}(e) + \Theta_{e}^{Po}(e),
\end{equation}
which is employed in Equation \ref{eq:reweighted convolution layer} to replace the former edge weight $\Theta_{e}^{Po}(e)$.

The newly introduced PRG module, by contrast, harnesses the powerful generative and reasoning capabilities of LLMs, along with their rich knowledge, as a crucial source of supplementary information. This information is seamlessly integrated into the representations of both games and players through Equations \ref{eq: MLP-embed-game} and \ref{eq: MLP-embed-player}, ensuring that both \textbf{global interests} and the \textbf{personal interests} are effectively \textbf{captured within the game and player representations}, respectively.

As depicted in Fig. \ref{fig:framework_tois}, the proposed framework, CPGRec+, consists of five core modules designed to deliver balance-oriented recommendations for players as follows:

\textbf{Accuracy-Driven Module (SGC+PER+\textcolor{black}{PRG})}: SGC module leverages cross-category associations to establish strict connections among games within the game graph, enhancing accuracy.

By employing a comparison method based on the Fisher distribution and information content, PER captures player preferences from historical interactions that significantly indicate the personal preferences of players. Interaction-wise edge weights are then applied to the player-game graph to refine player modeling. \textcolor{black}{Following a similar insight, PRG leverages LLMs to enrich player and game representations by incorporating complementary information, specifically, both the players' average ratings (represent global interests) and the dwelling time (represent personal interests). First, it prompts LLMs to generate game descriptions that encapsulate global interest. Subsequently, it constructs detailed player descriptions by inferring personal preferences through a comparative analysis of individual interest and global interest.}

\textbf{Diversity-Driven Module (CNA+PENR)}: In contrast to SGC, CNA emphasizes diverse-category neighbors to strengthen the connectivity of the game graph, promoting diversity. Unlike PER, PENR adjusts the weights of nodes and edges in the player-game graph based on game popularity, transforming popular games into conduits for disseminating messages from long-tail games, thereby enhancing diversity.

\textbf{Comprehensive Module}: It harmonizes accuracy and diversity by employing Combined Training with NSR. 

Together, these modules enable CPGRec+ to achieve a balance between accuracy and diversity in recommendations.

\subsection{Detailed Analysis of CPGRec+}
\label{Section: Validation of the Statistical Basis of PER Module}
In this section, we conduct a detailed analysis of CPGRec+, including both the validation of the standard normality of transformed dwelling time and average ratings involved in PER module and a case study for previously proposed PENR module.

\subsubsection{Validation of the Statistical Basis of PER Module.}
In this section, we employ the Kolmogorov-Smirnov Test for one sample to validate the efficacy of applying Box-Cox and Z-score transformation in mapping the dwelling time and average ratings to the standard normal distribution.

Kolmogorov-Smirnov Test (KS Test) for one sample estimates the likelihood that a given sample is drawn from a specified reference distribution. Specifically, the KS Test introduces the KS-statistic, formally defined as:
\begin{align}
    D_{n} &= sup_x |F_n(x)-F(x)|,
\end{align}
where $n$ is the size of sample, and 
\begin{align}
    F_n(x)=\frac{\# \text{ of sample points }\leq x}{n}
\end{align}
is the empirical cumulative distribution function, which measures the distance between the empirical distribution of the sample and that of the reference distribution, denoted as $F$. According to the Kolmogorov theorem, $\sqrt{n}D_n$ converges to the Kolmogorov distribution, whose cumulative distribution is given by  
\begin{align}
    F_{K}(x)=\frac{\sqrt{2\pi}}{x}\sum_{k=1}^{\infty}\exp(\frac{-(2k-1)^2\pi^2}{8x^2}),
\end{align}
which quantifies the probability that the given sample originates from the reference distribution. 

\begin{figure}[htbp]
  \begin{minipage}{0.35\textwidth}
    \centering
    \includegraphics[width=\textwidth]{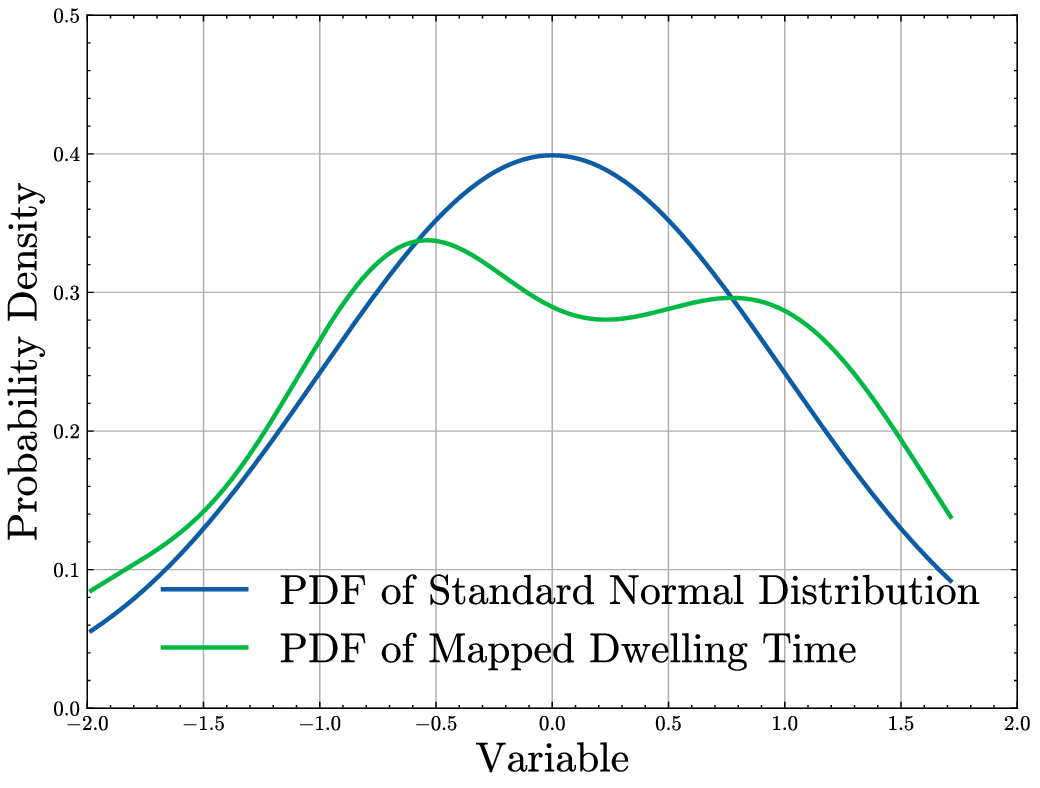}
  \end{minipage}%
  \hspace{1cm}
  \begin{minipage}{0.35\textwidth}
    \centering
    \includegraphics[width=\textwidth]{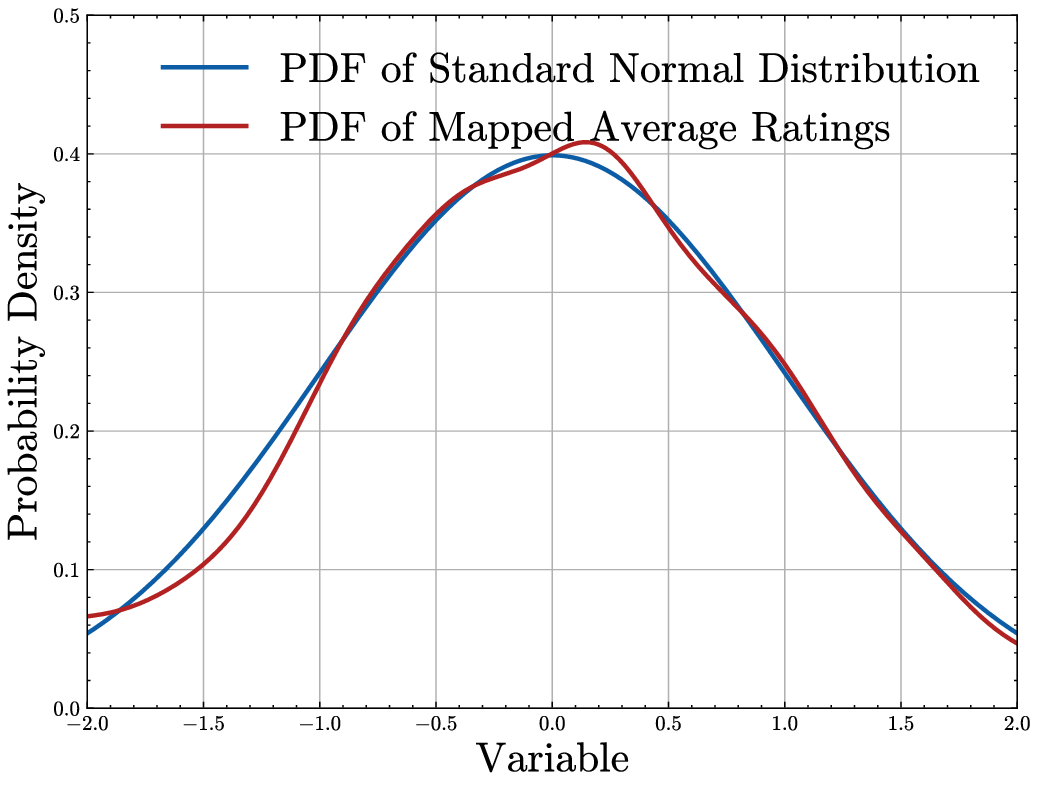}
  \end{minipage}
\vspace{-2mm}
\caption{Illustration of the probability density of dwelling time of players of \textit{Electronic Super Joy} and average ratings of all video games mapped by Box-Cox and Z-score transformations, which is intuitively highly close to standard normal distribution.}
\label{fig:valid_PER}
\end{figure}

In our analysis, the empirical sample consists of the dwelling times and average ratings transformed via Box-Cox and Z-score transformations, while the reference distribution is the standard normal distribution. The experimental results are shown in Fig. \ref{fig:valid_PER} and Table \ref{tab:valid_PER}. Among the 2,675 Steam video games with varying player dwelling times, we examine the game \textit{Electronic Super Joy} as a representative example.

As illustrated in Fig. \ref{fig:valid_PER}, the mapped dwelling times of players for \textit{Electronic Super Joy} and the average ratings of all games closely resemble the standard normal distribution in terms of probability density, providing an intuitive perspective on the effectiveness of the transformations. Furthermore, as shown in Table \ref{tab:valid_PER}, the p-values for dwelling time and average ratings are $0.5950$ and $0.5024$, respectively, both of which exceed the commonly used significance level of $\alpha=0.05$. This indicates that the \textbf{transformed data are statistically indistinguishable from the standard normal distribution}.

 \begin{table}[htbp]
 \caption{Experimental results of Kolmogorov-Smirnov test.}
\centering
\begin{tabular}{cccc}
\hline
Feature        & KS-statistic & P-value & Size of Sample \\ \hline
Dwelling Time  & 0.0861       & 0.5950  & 76        \\
Average Rating & 0.0159       & 0.5024  & 2675          \\ \hline
\end{tabular}
\label{tab:valid_PER}
\end{table}

\subsubsection{Case Study for PENR Module}
In our conference paper \cite{cpgrec}, we propose the PENR module, aimed at enhancing diversity by leveraging popular game nodes to propagate messages for long-tail games. This section includes a case study of the PENR module to provide a clearer and more intuitive understanding of its functionality. 

Consider a simplified scenario with only $2$ video games and $2$ players, denoted as $i_{0}, i_{1}$ and $u_{0}, u_{1}$, respectively. To show the role of PENR module in amplifying the influence of long-tail game nodes, we assume that $i_0$ is a long-tail game, while $i_1$ is popular. The existing historical interactions are $(u_{0},i_{0}), (u_{0},i_{1}), (u_{1},i_{1})$. Let \( e_{u_{0}}^{(l)}, e_{u_{1}}^{(l)}, e_{i_{0}}^{(l)}, e_{i_{1}}^{(l)} \) represent the representations of \( u_{0}, u_{1}, i_{0}, i_{1} \) respectively obtained from the \( l \)-th layer. According to the PENR module's definition, the operations of the graph convolutional layers are as follows:
\begin{align}
e_{u_1}^{(l+1)} &= C_{u_1u_1} \cdot e_{u_1}^{(l)} + e_h n_h \cdot C_{u_1i_1} \cdot e_{i_1}^{(l)}, \\
e_{i_1}^{(l+1)} &= n_h \cdot C_{i_1i_1} \cdot e_{i_1}^{(l)} + C_{u_1i_1} \cdot e_{u_1}^{(l)} + C_{u_0i_1} \cdot e_{u_0}^{(l)},\\
e_{u_0}^{(l+1)} &= n_l \cdot C_{u_0i_0} \cdot e_{i_0}^{(l)} + e_hn_hC_{u_0i_1} \cdot e_{i_1}^{(l)} + C_{u_0u_0} \cdot e_{u_0}^{(l)},\\
e_{i_0}^{(l+1)} &= n_l \cdot C_{i_0 i_0} \cdot e_{i_0}^{(l)} + C_{u_0 i_0} \cdot e_{u_0}^{(l)},
\end{align} \label{eq:pop weight}
where the normalization term is defined as:
\begin{align}
C_{a_mb_n} &= \frac{1}{\sqrt{|N_{a_m}|}\sqrt{|N_{b_n}|}}, \quad a, b \in \{u, i\}, m, n \in \{0, 1\},
\end{align}
and $e_h = \theta^{hot}_e, \quad n_h = \theta^{hot}_n, \quad n_l = \theta^{cold}_n$ are popularity-guided weights for edges and nodes. 
Consequently, the representation for  $u_{1}$ derived from the $3^{rd}$ layer could be obtained by recursion of Equation \ref{eq:pop weight}, formally denoted as:
\begin{align}
    e_{u_1}^{(l+3)} = A e_{i_0}^{(l)} + B e_{i_1}^{(l)} + C e_{u_0}^{(l)} + D e_{u_1}^{(l)},
\end{align}
where
\begin{align}
    A &= e_hn_hn_lC_{u_1i_1}C_{u_0i_1}C_{u_0i_0},\\
    B &= e_hn_hC_{u_1u_1}^{2}C_{u_1i_1}+e_hn_h^{2}C_{u_1i_1}C_{i_1i_1}C_{u_1u_1},\\
    &+e_hn_h^{3}C_{u_1i_1}C_{i_1i_1}^{2}+e_h^{2}n_h^{2}C_{u_1i_1}^{3}+e_h^{2}n_h^{2}C_{u_0i_1}^{2}C_{u_1i_1}\\
    C &= e_hn_hC_{u_1i_1}C_{u_0i_1}C_{u_1u_1}+e_hn_h^{2}C_{u_1i_1}C_{i_1i_1}C_{u_0i_1}\\
    &+e_hn_hC_{u_1i_1}C_{u_0i_1}C_{u_0u_0},\\
    D &= C_{u_1u_1}^3 + e_hn_hC_{u_1u_1}C_{u_1i_1}^2\\
    &+ e_hn_h^2C_{u_1i_1}^2C_{i_1i_1}+e_hn_hC_{u_1i_1}^2C_{u_1u_1}.
\end{align}

To explore the function of the PENR module in enhancing the influences of long-tail games, we define the influence index $INF$ to quantify the influence of each node on $u_1$:
\begin{align}
    INF_{i_0} &= \frac{A}{A+B+C+D},\\
    INF_{i_1} &= \frac{B}{A+B+C+D},\\
    INF_{u_0} &= \frac{C}{A+B+C+D},\\
    INF_{u_1} &= \frac{D}{A+B+C+D}.\\
\end{align} \label{eq:INF def}
\begin{table}[h]
\centering
\caption{Influence factors.}
\begin{tabular}{cccccc}
\hline
$e_h/n_h/n_l$ & $INF_{i_{0}}$ & $INF_{i_{1}}$ & $INF_{u_{0}}$ & $INF_{u_{1}}$ \\ \hline
1/1/1 & 0.0747 & 0.3796 & 0.1966 & 0.3491 \\ 
1/1/6 & 0.3263 & 0.2764 & 0.1432 & 0.2542 \\
1/0.2/1 & 0.0864 & 0.2549 & 0.1922 & 0.4665 \\
5/1/1 & 0.0486 & 0.6439 & 0.1280 & 0.1796 \\
5/0.2/1 & 0.0966 & 0.3379 & 0.1912 & 0.3742 \\
5/0.2/6 & 0.3908 & 0.2279 & 0.1290 & 0.2523 \\
\hline
\end{tabular}
\label{tab:INF}
\end{table}

Based on Equation \ref{eq:INF def}, we conduct the following case study as shown in Table \ref{tab:INF} by varying the popularity-guided weights $n_h, n_l, e_h$. It is worth noting that $n_h=n_l=e_h=1$ means that the PENR module is removed. From the results, there are several observations:
\begin{itemize}
    \item  \textbf{Effectiveness of Popularity-guided Reweighting:} Compared to the scenario without PENR module (\textit{i.e.} $e_h/n_h/n_l=1/1/1$), the $INF_{i_0}$ with PENR module (\textit{i.e.} $e_h/n_h/n_l=5/0.2/6$) significantly increases from $0.0747$ to $0.3908$, which proves the effectiveness of popularity-guided reweighting. 
    \item \textbf{Increase of Long-Tail Game Node Weight:}  The setup $e_h/n_h/n_l=1/1/6$ enlarges the effectiveness of long-tail game node $i_0$ by amplifying its node weight,  resulting in an increase in the influence index $INF_{i_0}$ to $0.3263$, while concurrently reducing the influence index of the popular game node $i_1$ by approximately $27.19\%$.
    \item \textbf{Decrease of Popular Game Node Weight:} The setup $e_h/n_h/n_l=1/0.2/1$ reduces the node weight of the popular game $i_1$, thereby significantly constraining its influence; consequently, $INF_{i_1}$ decreases by about $32.85\%$, while the influence index of all other nodes increase.
    \item  \textbf{Increase of Popular Game Edge Weight:} The setup $e_h/n_h/n_l=5/1/1$ enlarges the weight of outgoing edges from the popular game node $i_1$, thereby enhancing its capacity to propagate messages. This transformation leads to an expansion of the influence of the popular game $i_1$. 
    
    \item  \textbf{Combined Effect of Increasing Popular Game Edge Weight and Decreasing Popular Game Node Weight:} The setup $e_h/n_h/n_l=5/0.2/1$ further reduces $n_h$ compared to the last setup, thereby limiting the effectiveness of $i_1$, and resulting in a decrease in $INF_{i_1}$ by $47.52\%$.  
\end{itemize}

The aforementioned examples intuitively illustrate and substantiate the importance of the popularity-guided weight introduced by the PENR module. Together, these mechanisms enhance the influence of long-tail games while simultaneously mitigating the predominance of popular games.

\subsubsection{Complexity Analysis of CPGRec+}

In this section, we meticulously analyze the complexity of CPGRec+ to foster the potential real-world deployment of the framework. This includes not only the fine-grained complexity analysis of the multiple components but also the comparison between CPGRec+ and other state-of-the-art recommender systems.

\textbf{Complexity Analysis of Stringency-improved Game Connection (SGC)}. As the first accuracy-driven module, SGC includes three essential steps: (1) constructing game graphs with stringency-improved game connections $\mathcal{G}^{g\&d},\mathcal{G}^{g\&p},\mathcal{G}^{d\&p}$, (2) game representation learning on $\mathcal{G}^{g\&d},\mathcal{G}^{g\&p},\mathcal{G}^{d\&p}$ with the LightGCN model, (3) graph attention across multiple graphs.

The following is a detailed complexity analysis of each of the above steps. Since the \textbf{first} step is completed before model inference, it will not be taken into account in the time complexity analysis. In the \textbf{second} step, we use the LightGCN model to learn node representations on each constructed strict graph. Since LightGCN performs one embedding propagation step, its time complexity is $O(d_{game}|E^{strict}|)$, where $|E^{g\&d}|,|E^{g\&p}|,|E^{d\&p}|$ are the number of edges in the graphs $\mathcal{G}^{g\&d},\mathcal{G}^{g\&p},\mathcal{G}^{d\&p}$, respectively, $|E^{strict}|$ is the summation of $|E^{g\&d}|,|E^{g\&p}|,|E^{d\&p}|$ and $d_{game}$ is the feature dimension of the game nodes. Specifically, this process involves an Embedding Propagation step and a Layer Combination step, with time complexities of $O(d_{game}|E^{strict}|)$ and $O(d_{game}|\mathcal{I}|)$ respectively. Since $|E^{strict}|$ is usually greater than $|\mathcal{I}|$, the total time complexity of this step is $O(d_{game}|E^{strict}|)$. In the \textbf{third} step, we apply Graph Attention Network (GAT) across the strict graphs $\mathcal{G}^{g\&d},\mathcal{G}^{g\&p},\mathcal{G}^{d\&p}$ to fuse the learned representations. The total time complexity of this step is $O(|\mathcal{I}|{d_{game}}^2+|E^{strict}|d_{game})$. This can be attributed to the fact that GAT involves linear mapping for each of $|\mathcal{I}|$ nodes and computing attention scores for each of $|E^{strict}|$ edges, with time complexities of $O(|\mathcal{I}|{d_{game}}^2)$ and $O(|E^{strict}|d_{game})$, respectively. As a result, the complexity of the SGC component is $O(|\mathcal{I}|{d_{game}}^2+|E^{strict}|d_{game})$.

\textbf{Complexity Analysis of Connectivity-enhanced Neighbor Aggregation (CNA)}. CNA is the first diversity-driven module. Highly similar to SGC, it consists of three key steps: (1) constructing a connectivity-enhancecd game graph $\mathcal{G}^{Co}$, (2) game representation learning on $\mathcal{G}^{Co}$ with $k$-layers of LightGCN, (3) graph attention across multiple layers. Considering that the only difference is the number of layers of LightGCN used, by following the aforementioned detailed analysis process as follows, the complexity of CNA is $O(|\mathcal{I}|{d_{game}}^2+k|E^{Co}|d_{game})$, where $k$ and $|E^{Co}|$ are the number of the used LightGCN layers and that of edges within graph $\mathcal{G}^{Co}$, respectively.

\textbf{Complexity Analysis of Preference-informed Edge Reweighting (PER)}. PER is the second accuracy-driven module as well as one of the newly proposed extensions. It involves two steps: (1) Fisher Distribution-based Sign Decision, (2) Information Content-based Volume Evaluation. Based on the dwelling time and the average rating data from the dataset, PER models a player's personal preferences by assigning signed weights to historical player-game interactions, which are determined by comparing personal interest (dwelling time) against global interest (average game ratings) to detect both the significant interest and disinterest. It is worth highlighting the computational efficiency of the PER method. Since the calculation of its edge weights relies solely on static historical data, namely dwelling times and average ratings, these weights can be entirely pre-computed offline and remain constant during model inference.

\textbf{Complexity Analysis of Preference-informed Representation Generation (PRG)}. PRG is the third accuracy-driven module and a newly proposed extention in this study, including (1) Rating-informed Game Description Generation, (2) Preference-informed Player Description Generation, and (3) Representation Embedding and Integration. Conceptually, the process begins with constructing a prompt for each game and player, after which an LLM is invoked to generate their textual descriptions. Finally, a large embedding model converts these descriptions into numerical representations, which are subsequently fused with the existing node representations via MLP models. Notably, the representation embedding involved in steps (1) and (2), as well as partly in step (3), can be completed in advance; only the semantic alignment and fusion with MLP models, as shown in Equations \ref{eq: MLP-embed-game} and \ref{eq: MLP-embed-player}, must be performed at inference time. This design significantly improves the recommendation efficiency of CPGRec+. Specifically, the complexity of alignment step with $MLP^{\mathcal{D}-Align}$ is $O((|\mathcal{I}|+|\mathcal{U}|)\cdot(d_{emb}d_h+d_h d_{shared}))$, respectively, where $d_{emb}$ is the embedding dimension of the adopted embedding model (\textit{i.e.,} of M3 in this study), $d_h$ and $d_{shared}$ are the dimensions of output of the hidden layer and output layer, respectively. Furthermore, since $d_{emb}\gg d_{h}$, the complexity of alignment methods can be simplified to $O((|\mathcal{I}|+|\mathcal{U}|)\cdot (d_{emb}d_h))$. Similarly, the complexity of the integration step is $O((|\mathcal{I}|+|\mathcal{U}|)\cdot (d_hd_{shared}))$. As a result, taking the aforementioned two steps into account, the total complexity of PRG component is $O((|\mathcal{I}|+|\mathcal{U}|)\cdot (d_{embed}d_h))$ considering $d_{emb}\gg d_{shared}$.

\textbf{Complexity Analysis of Popularity-guided Edges and Nodes Reweighting (PENR)}. As the second diversity-driven module, PENR enhances long-tail game exposure by reweighting nodes and edges in the player-game bipartite graph based on popularity. The PENR weights are pre-computed offline from static popularity data and thus do not add to the inference time complexity. During inference, these weights are combined with PER weights and integrated into the LightGCN message-passing mechanism (Equation \ref{eq:rwt combination}). This efficient application, an element-wise multiplication, does not increase the asymptotic complexity of the graph convolution. The overall complexity is therefore $O(|E^{bi}|d_{shared})$, where $|E^{bi}|$ is the number of bipartite edges and $d_{shared}$ is the representation dimensionality.

\textbf{Overall Complexity of CPGRec+.} In summary, the total inference time complexity of CPGRec+ is the summation of its online components: $O(|\mathcal{I}|{d_{game}}^2 + (|E^{strict}|+k|E^{Co}|)d_{game} + (|\mathcal{I}|+|\mathcal{U}|)(d_{emb}d_h) + |E^{bi}|d_{shared})$. In practical recommender system scenarios, the number of user-item interactions, $|E^{bi}|$, is typically several orders of magnitude larger than the number of item-item connections ($|E^{strict}|, |E^{Co}|$) and the total number of nodes ($|\mathcal{I}|+|\mathcal{U}|$). This leads to the following three key observations. 
\begin{itemize}
    \item \textbf{The PENR component is the primary computational bottleneck of CPGRec+}. This can be attributed to the fact that the complexity is dominated by the graph convolution on the player-game bipartite graph leveraged by PENR.
    \item \textbf{Newly introduced PER and PRG components are not only pivotal but also efficient components of CPGRec+}. They significantly enhance the balance-oriented recommendation capabilities without substantially increasing the framework's online computational load. 
    \item \textbf{CPGRec+ is a highly deployable framework achieving a favorable balance between performance and complexity}. While achieving superior performance, CPGRec+ exhibits a complexity that is entirely on par with a wide range of state-of-the-art GNN-based recommender systems \cite{he2020lightgcn, Yang_2022, Zheng_2021, Yang_2023}, surpassed only by efficient the MVGNN \cite{mvgnn} model that circumvents direct processing of the entire graph structure through meticulous designs. The comparison results are organized in Table \ref{tab: complexity comparison}. This is attributed to the extensive reliance on the user-item graph among these broad GNN-based recommender systems, which also serves as the primary computational bottleneck for CPGRec+. Furthermore, CPGRec+ demonstrates even greater deployability compared to methods such as SURGE and DGRec, which depend on the dynamic construction of sequences or subgraphs during inference. Besides, in terms of designs, CPGRec+ also guarantees exceptional inference efficiency through its broad incorporation of lightweight LightGCN layers and a substantial volume of critical pre-computed weights.
\end{itemize}

\begin{table*}
\caption{Complexity comparison between our proposed CPGRec+ and state-of-the-art GNN-based recommender systems. In the comparison, $\checkmark, \times$, and $=$ indicate that the complexity of the model is lower than, higher than, and equal to CPGRec+, respectively.}
\centering
\begin{adjustbox}{max width=\textwidth}
    \begin{tabular}{@{}ccccccc@{}}
\toprule
Model & CPGRec+ & SCGRec & DGCN & DGRec & MVGNN & LightGCN \\ \midrule
Graph Convolution on Bipartite Graph & yes & yes & yes & yes & no & yes \\
Complexity Comparison w/ CPGRec+ & $=$ & $\times$ & $\times$ & $\times$ & $\checkmark$ & $\checkmark$ \\ \bottomrule
\end{tabular}
\end{adjustbox}\label{tab: complexity comparison}
\end{table*}

\section{Experiments}
In this section, we conduct comprehensive experiments to evaluate the performance of CPGRec+, focusing on its accuracy and diversity.

\subsection{Experimental Setup}
\subsubsection{Dataset}


\textcolor{black}{As Steam is one of the largest video-game distribution platforms globally, its specialized datasets are notably representative, readily accessible, and have been extensively utilized in prior research. This motivates our use of two real-world Steam datasets as the empirical foundation for our experiments: (1) The first, which we denote as \textbf{Steam I}, is the dataset previously employed in CPGRec \cite{cpgrec} and other established works \cite{o2016condensing, Yang_2022}; (2) following existing works \cite{steam2_1, steam2_2}, we further introduce a different version of steam data, designated as \textbf{Steam II}, which encompasses player-game interaction data spanning from October 2010 to January 2018, as well as a rich array of supplementary information like dwelling time, pricing details, average ratings, and game categorical information. Table \ref{table:dataset} succinctly presents the basic statistics of Steam I and Steam II datasets, offering a clear overview of their features.}

\begin{table}[htbp]
\centering
\caption{Statistics of the Steam I and II datasets.}
\begin{tabular}{@{}cccccc@{}}
\toprule
Dataset & & Steam I & & Steam II & \\ \cmidrule(r){1-6}
\# Players & & 3,908,744 & & 334,730 & \\
\# Games & & 2,675 & & 13,047& \\
\# Interactions & & 95,208,806 & & 3,686,172& \\ \cmidrule(r){1-6}
Genre Information & & $\checkmark$ & & $\checkmark$& \\
Developer Information & & $\checkmark$ & & $\checkmark$& \\
Publisher Information & & $\checkmark$ & & $\checkmark$& \\ \cmidrule(r){1-6}
Dwelling Time & & $\checkmark$ & & $\checkmark$& \\ 
Average Rating & & $\checkmark$ & & $\checkmark$& \\ 
\bottomrule
\end{tabular}
\label{table:dataset}
\end{table}

For Steam I, in accordance with the partition setting of previous studies \cite{cpgrec}, the observed historical interactions are split into train, valid, and test sets, following a ratio of $0.8/0.1/0.1$; for Steam II, the splitting ratio is $0.5/0.05/0.45$.

\subsubsection{Evaluation Metrics}
For accuracy assessment, we employ \textit{NDCG@K, Recall@K, Hit@K}, and \textit{Precision@K} as our metrics; for diversity, in addition to \textit{Coverage@K} and \textit{Entropy@K}, which were employed in our previous conference work \cite{cpgrec}, we further introduce \textit{Conventional Coverage@K}, \textcolor{black}{\textit{Tail Coverage@K}, and \textit{Tail@K}} defined as:
\begin{align}
    \text{\textit{Conventional Coverage}}@K &= \frac{|\cup_{u\in \mathcal{U}}\mathcal{I}_K^{(u)}|}{|\mathcal{I}|},\\
    \text{\textit{Tail Coverage}}@K &= \frac{|\cup_{u\in \mathcal{U}}(\mathcal{I}_K^{(u)}\cap \mathcal{I}^{LT})|}{|\mathcal{I}^{LT}|},\\
    \text{\textit{Tail}}@K &= \frac{1}{|\mathcal{U}|}\sum_{u\in\mathcal{U}}\frac{\mathcal{I}_K^{(u)}\cap \mathcal{I}^{LT}}{K},
\end{align}
where $\mathcal{I}_K^{(u)}$ is the top-K recommended list for each user $u\in \mathcal{U}$, and $K$ takes values in $\{5,10\}$, and \textcolor{black}{$\mathcal{I}^{LT}$ is the set of long-tail games. More specifically, they are defined as the 20\% of games with the lowest player counts in this study.} It is crucial to note that \textit{Coverage@K} and \textit{Entropy@K} evaluate individual diversity, whereas \textit{Conventional Coverage@K} assesses global diversity. Specifically, \textit{Coverage@K} measures diversity across genres, developers, publishers, and their combined total in the top-$K$ recommendations; \textit{Entropy@K} evaluates the entropy of recommendation lists based on these categories; \textit{Conventional Coverage@K} assesses the variety of unique video games present in the top-K recommendations.

\subsection{Baselines}
In order to thoroughly assess the performance of CPGRec+, we carried out comparisons against a variety of established recommender systems.


\textcolor{black}{In terms of \textbf{accuracy}, we compared CPGRec+ with five models: the widely-used GNN-based model LightGCN \cite{he2020lightgcn}, and the state-of-the-art recommender systems SURGE\cite{surge}, SCGRec \cite{Yang_2022}, MVGNN \cite{mvgnn}, and BIGCF \cite{bigcf}, which leads accuracy-oriented recommender systems.}

In terms of\textbf{diversity}, CPGRec+ is evaluated in comparison to five distinguished diversity-focused models: MMR \cite{carbonell1998use}, EDUA \cite{liang2021enhancing}, DDGraph \cite{10.1145/3460231.3478845}, DGCN \cite{Zheng_2021}, and DGRec \cite{Yang_2023}. 

In terms of \textbf{balance}, EXPLORE \cite{explore} and the prior work CPGRec\cite{cpgrec} were introduced as trade-off-focused baselines.

\subsection{Performance Evaluation}

\begin{table*}[htbp]
\centering
\caption{Accuracy-focus performance comparison of different accuracy-, diversity- and balance-driven recommender systems in terms of \textit{NDGC, Recall, Hit, Precision} on Steam I. }
\begin{adjustbox}{max width = 0.8\textwidth} 
\begin{tabular}{@{}clcccccccccccccccccccccccccccccccccc@{}}
\toprule
\multicolumn{2}{c}{\multirow{2}{*}{}} & & \multirow{2}{*}{Methods} & & \multicolumn{2}{c}{NDGC} & & \multicolumn{2}{c}{Recall} & & \multicolumn{2}{c}{Hit} & & \multicolumn{2}{c}{Precision} \\
\cmidrule(lr){6-7} \cmidrule(lr){9-10} \cmidrule(lr){12-13} \cmidrule(lr){15-16}
\multicolumn{2}{c}{} & & & & @5 & @10 & & @5 & @10 & & @5 & @10 & & @5 & @10 \\
\midrule
\multicolumn{2}{c}{\multirow{6}{*}{\begin{tabular}[c]{@{}c@{}}Accuracy-driven\\ Methods\end{tabular}}} & & LightGCN & & 0.1861 & 0.2100 & & 0.2452 & 0.3174 & & 0.2849 & 0.3784 & & 0.0637 & 0.0447 \\
\multicolumn{2}{c}{} & & SURGE & & 0.3133 & 0.3521 & & 0.4128 & 0.5340 & & 0.4779 & 0.6361 & & 0.1065 & 0.0753 \\
\multicolumn{2}{c}{} & & BIGCF & & 0.3986 & 0.4271 & & 0.5017 & 0.6053 & & 0.6184 & 0.7296 & & 0.1391 & 0.0915 \\
\multicolumn{2}{c}{} & & MVGNN & & 0.4138 & 0.4452 & & 0.4954 & 0.5916 & & 0.6129 & 0.7183 & & 0.1446 & 0.0957 \\
\multicolumn{2}{c}{} & & SCGRec & & 0.4351 & 0.4660 & & 0.5385 & 0.6311 & & 0.6519 & 0.7535 & & 0.1508 & 0.0969 \\
\multicolumn{2}{c}{} & & Accuracy-focused CPGRec & & \underline{0.4796} & \underline{0.5000} & & \underline{0.5746} & \underline{0.6387} & & \underline{0.6983} & \underline{0.7659} & & \underline{0.1625} & \underline{0.0989} \\
\multicolumn{2}{c}{} & & Accuracy-focused CPGRec+ & & \textbf{0.4805} & \textbf{0.5029} & & \textbf{0.5768} & \textbf{0.6413} & & \textbf{0.6993} & \textbf{0.7674} & & \textbf{0.1637} & \textbf{0.0993} \\
\midrule
\multicolumn{2}{c}{\multirow{7}{*}{\begin{tabular}[c]{@{}c@{}}Diversity-driven\\ Methods\end{tabular}}} & & MMR & & 0.3259 & 0.3871 & & 0.2768 & 0.3420 & & 0.3522 & 0.4302 & & 0.0684 & 0.0623 \\
\multicolumn{2}{c}{} & & EDUA & & 0.3839 & 0.4072 & & 0.4746 & 0.5545 & & 0.5808 & 0.6936 & & 0.1149 & 0.0826 \\
\multicolumn{2}{c}{} & & DDGraph & & 0.3997 & 0.4298 & & 0.4949 & 0.5883 & & 0.6059 & 0.7094 & & 0.1399 & 0.0894 \\
\multicolumn{2}{c}{} & & DGCN & & 0.3732 & 0.4025 & & 0.4536 & 0.5523 & & 0.6056 & 0.7123 & & 0.1129 & 0.0806 \\
\multicolumn{2}{c}{} & & DGRec & & 0.3546 & 0.3982 & & 0.4293 & 0.5434 & & 0.5791 & 0.7126 & & 0.1041 & 0.0786 \\
\multicolumn{2}{c}{} & & Diversity-focused CPGRec & & \underline{0.4285} & \underline{0.4547} & & \underline{0.5168} & \underline{0.5990} & & \underline{0.6390} & \underline{0.7292} & & \underline{0.1469} & \underline{0.0922} \\
\multicolumn{2}{c}{} & & Diversity-focused CPGRec+ & & \textbf{0.4312} & \textbf{0.4570} & & \textbf{0.5221} & \textbf{0.6029} & & \textbf{0.6439} & \textbf{0.7322} & & \textbf{0.1485} & \textbf{0.0930} \\
\midrule
\multicolumn{2}{c}{\multirow{3}{*}{\begin{tabular}[c]{@{}c@{}}Balance-driven\\ Methods\end{tabular}}} & & EXPLORE & & 0.4126 & 0.4381 & & 0.5002 & 0.5895 & & 0.6187 & 0.7177 & & 0.1398 & 0.0916 \\
\multicolumn{2}{c}{} & & CPGRec (trade-off framework) & & \underline{0.4320} & \underline{0.4547} & & \underline{0.5168} & \underline{0.5990} & & \underline{0.6390} & \underline{0.7292} & & \underline{0.1469} & \underline{0.0922} \\
\multicolumn{2}{c}{} & & CPGRec+ (trade-off framework) & & \textbf{0.4422} & \textbf{0.4691} & & \textbf{0.5347} & \textbf{0.6192} & & \textbf{0.6590} & \textbf{0.7473} & & \textbf{0.1538} & \textbf{0.0967} \\
\bottomrule
\end{tabular}
\end{adjustbox}
\label{table: performance comparison with new metric: accu and cc on steam I}
\end{table*}

\begin{table*}[htbp] 
\centering
 \caption{Accuracy-focus performance comparison of different accuracy-, diversity- and balance-driven recommender systems in terms of \textit{NDGC, Recall, Hit, Precision} on Steam II. } 
 \begin{adjustbox}{max width = 0.8\textwidth} 
 \begin{tabular}{@{}clcccccccccccccccccccccccccccccccccc@{}} 
 \toprule 
 \multicolumn{2}{c}{\multirow{2}{*}{}} & & \multirow{2}{*}{Methods} & & \multicolumn{2}{c}{NDGC} & & \multicolumn{2}{c}{Recall} & & \multicolumn{2}{c}{Hit} & & \multicolumn{2}{c}{Precision} \\ 
 \cmidrule(lr){6-7} \cmidrule(lr){9-10} \cmidrule(lr){12-13} \cmidrule(lr){15-16} 
 \multicolumn{2}{c}{} & & & & @5 & @10 & & @5 & @10 & & @5 & @10 & & @5 & @10 \\ 
 \midrule 
 \multicolumn{2}{c}{\multirow{6}{*}{\begin{tabular}[c]{@{}c@{}}Accuracy-driven\\ Methods\end{tabular}}} & & LightGCN & & 0.0505 & 0.0579 & & 0.0396 & 0.0625 & & 0.1765 & 0.2566 & & 0.0409 & 0.0324 \\ 
 \multicolumn{2}{c}{} & & SURGE & & 0.1913 & 0.2287 & & 0.1804 & 0.2691 & & 0.5926 & 0.7438 & & 0.1612 & 0.1229 \\ 
 \multicolumn{2}{c}{} & & BIGCF & & 0.2403 & 0.2886 & & 0.2267 & 0.3431 & & 0.6782 & 0.8459 & & 0.1991 & 0.1554 \\ 
 \multicolumn{2}{c}{} & & MVGNN & & 0.2427 & 0.2902 & & 0.2281 & 0.3467 & & 0.6813 & 0.8492 & & 0.2008 & 0.1571 \\ 
 \multicolumn{2}{c}{} & & SCGRec & & 0.2391 & 0.2864 & & 0.2243 & 0.3408 & & 0.6754 & 0.8416 & & 0.1973 & 0.1539 \\ 
 \multicolumn{2}{c}{} & & Accuracy-focused CPGRec & & \underline{0.2769} & \underline{0.3341} & & \underline{0.2638} & \underline{0.3953} & & \underline{0.7257} & \underline{0.8834} & & \underline{0.2116} & \underline{0.1642} \\ 
 \multicolumn{2}{c}{} & & Accuracy-focused CPGRec+ & & \textbf{0.2794} & \textbf{0.3368} & & \textbf{0.2662} & \textbf{0.3989} & & \textbf{0.7288} & \textbf{0.8871} & & \textbf{0.2139} & \textbf{0.1658} \\ 
 \midrule 
 \multicolumn{2}{c}{\multirow{7}{*}{\begin{tabular}[c]{@{}c@{}}Diversity-driven\\ Methods\end{tabular}}} & & MMR & & 0.1878 & 0.2246 & & 0.1763 & 0.2659 & & 0.5831 & 0.7324 & & 0.1581 & 0.1207 \\ 
 \multicolumn{2}{c}{} & & EDUA & & 0.2216 & 0.2673 & & 0.2088 & 0.3124 & & 0.6509 & 0.8172 & & 0.1923 & 0.1506 \\ 
 \multicolumn{2}{c}{} & & DDGraph & & 0.2184 & 0.2619 & & 0.2046 & 0.3065 & & 0.6432 & 0.8058 & & 0.1874 & 0.1469 \\ 
 \multicolumn{2}{c}{} & & DGCN & & 0.2247 & 0.2708 & & 0.2113 & 0.3169 & & 0.6561 & 0.8214 & & 0.1948 & 0.1522 \\ 
 \multicolumn{2}{c}{} & & DGRec & & 0.2031 & 0.2492 & & 0.1917 & 0.2863 & & 0.6188 & 0.7783 & & 0.1746 & 0.1341 \\ 
 \multicolumn{2}{c}{} & & Diversity-focused CPGRec & & \underline{0.2486} & \underline{0.2982} & & \underline{0.2354} & \underline{0.3547} & & \underline{0.6917} & \underline{0.8583} & & \underline{0.2039} & \underline{0.1601} \\ 
 \multicolumn{2}{c}{} & & Diversity-focused CPGRec+ & & \textbf{0.2512} & \textbf{0.3013} & & \textbf{0.2381} & \textbf{0.3586} & & \textbf{0.6954} & \textbf{0.8621} & & \textbf{0.2064} & \textbf{0.1613} \\ 
 \midrule 
 \multicolumn{2}{c}{\multirow{3}{*}{\begin{tabular}[c]{@{}c@{}}Balance-driven\\ Methods\end{tabular}}} & & EXPLORE & & 0.2352 & 0.2881 & & 0.2239 & 0.3426 & & 0.6723 & 0.8394 & & 0.1981 & 0.1557 \\ 
 \multicolumn{2}{c}{} & & CPGRec (trade-off framework) & & \underline{0.2819} & \underline{0.3398} & & \underline{0.2688} & \underline{0.4029} & & \underline{0.7319} & \underline{0.8887} & & \underline{0.2148} & \underline{0.1667} \\ 
 \multicolumn{2}{c}{} & & CPGRec+ (trade-off framework) & & \textbf{0.2844} & \textbf{0.3427} & & \textbf{0.2709} & \textbf{0.4052} & & \textbf{0.7349} & \textbf{0.8920} & & \textbf{0.2162} & \textbf{0.1679} \\ 
 \bottomrule 
 \end{tabular} 
 \end{adjustbox} 
 \label{table: performance comparison with new metric: accu and cc on steam II} 
 \end{table*}

\begin{table*}[htbp]
\centering
\caption{Diversity-focus performance comparison of different accuracy-, diversity- and balance-driven recommender systems in terms of \textit{Conventional Coverage (CC), Coverage (C)} and \textit{Entropy (E)} on Steam I.}
\begin{adjustbox}{max width = \textwidth}
\begin{tabular}{@{}clcccccccccccccccccccccccccccccccccc@{}}
\toprule
\multicolumn{2}{c}{\multirow{2}{*}{}} & & \multirow{2}{*}{Methods} & & \multicolumn{2}{c}{CC} & & \multicolumn{2}{c}{C(genre)} & & \multicolumn{2}{c}{C(developer)} & & \multicolumn{2}{c}{C(publisher)} & & \multicolumn{2}{c}{C(total)} & & \multicolumn{2}{c}{E(genre)} & & \multicolumn{2}{c}{E(developer)} & & \multicolumn{2}{c}{E(publisher)} \\
\cmidrule(lr){6-7} \cmidrule(lr){9-10} \cmidrule(lr){12-13} \cmidrule(lr){15-16} \cmidrule(lr){18-19} \cmidrule(lr){21-22} \cmidrule(lr){24-25} \cmidrule(lr){27-28}
\multicolumn{2}{c}{} & & & & @5 & @10 & & @5 & @10 & & @5 & @10 & & @5 & @10 & & @5 & @10 & & @5 & @10 & & @5 & @10 & & @5 & @ 10\\
\midrule
\multicolumn{2}{c}{\multirow{7}{*}{\begin{tabular}[c]{@{}c@{}}Accuracy-driven\\ Methods\end{tabular}}} & & LightGCN & & 0.2373 & 0.2848 & & 2.2839 & 3.2916 & & 3.0915 & 5.1401 & & 2.2818 & {4.3852} & & 7.6572 & 12.8169 & & 0.9409 & 0.9011 & & 0.7030 & 1.4600 & & 1.4820 & 1.2652 \\
\multicolumn{2}{c}{} & & SURGE & & 0.2247 & 0.2718 & & 1.4131 & 2.0649 & & 3.0593 & 5.1238 & & 2.0577 & 3.8481 & & 7.2351 & 12.0368 & & 0.8519 & 0.8184 & & 0.6538 & 1.3815 & & 1.3581 & 1.1949 \\
\multicolumn{2}{c}{} & & BIGCF & & 0.2491 & 0.2908 & & 2.3142 & 3.3611 & & 3.1105 & 5.2582 & & 2.3088 & 4.4951 & & 7.7335 & 13.1144 & & 0.9693 & 0.9317 & & 0.7149 & 1.5217 & & 1.4992 & 1.3038 \\
\multicolumn{2}{c}{} & & MVGNN & & 0.2536 & 0.2949 & & 2.4218 & 3.5381 & & 3.1672 & 5.3991 & & 2.3754 & 4.6942 & & 7.9644 & 13.6314 & & 0.9601 & 0.9255 & & 0.7103 & 1.4892 & & 1.5134 & 1.2916 \\
\multicolumn{2}{c}{} & & SCGRec & & {0.2416} & {0.2785} & & {2.3871} & {3.3754} & & \textbf{3.1898} & {5.4122} & & {2.3708} & 4.1812 & & {7.9477} & {12.9688} & & {0.9766} & {0.9118} & & {0.7297} & {1.4773} & & {1.5383} & {1.2802} \\
\multicolumn{2}{c}{} & & Accuracy-focused CPGRec & & \underline{0.2685} & \underline{0.3188} & & \textbf{2.7509} & \textbf{4.3680} & & \underline{3.1800} & \textbf{6.2738} & & \textbf{2.5783} & \textbf{4.8981} & & \textbf{8.5092} & \textbf{15.5399} & & \textbf{1.0455} & \textbf{1.0926} & & \textbf{0.7812} & \textbf{1.7702} & & \textbf{1.6470} & \textbf{1.5340} \\
\multicolumn{2}{c}{} & & Accuracy-focused CPGRec+ & & \textbf{0.2757} & \textbf{0.3291} & & \underline{2.7491} & \underline{4.3640} & & {3.1768} & \underline{6.2719} & & \underline{2.5674} & \underline{4.8931} & & \underline{8.4933} & \underline{15.5290} & & \underline{1.0404} & \underline{1.0899} & & \underline{0.7692} & \underline{1.7685} & & \underline{1.6431} & \underline{1.5328} \\
\midrule
\multicolumn{2}{c}{\multirow{7}{*}{\begin{tabular}[c]{@{}c@{}}Diversity-driven\\ Methods\end{tabular}}} & & MMR & & 0.2719 & {0.3284} & & {2.5928} & 3.8183 & & 3.5418 & 5.4192 & & {3.0945} & {4.8985} & & 9.2291 & {14.1360} & & {1.0783} & {1.0708} & & 0.7778 & {1.7142} & & 1.6190 & {1.4657} \\
\multicolumn{2}{c}{} & & EDUA & & 0.2496 & 0.2925 & & 2.1285 & 3.2491 & & 3.2595 & 5.4325 & & 2.3465 & 5.2056 & & 7.7345 & 13.8872 & & 0.9900 & 0.9821 & & 0.7077 & 1.5624 & & 1.4660 & 1.3385 \\
\multicolumn{2}{c}{} & & DDGraph & & 0.2871 & 0.3033 & & 2.3905 & 3.7699 & & 3.0540 & 5.4324 & & 2.6843 & 4.6402 & & 8.1288 & 13.8425 & & 0.9956 & 0.8848 & & {0.8078} & 1.4931 & & 1.5679 & 1.4212 \\
\multicolumn{2}{c}{} & & DGCN & & 0.2602 & 0.3090 & & 2.4069 & 4.2905 & & 3.2863 & 5.8637 & & 2.4981 & 3.9211 & & 8.1913 & 14.0753 & & 1.0133 & 1.0086 & & 0.7301 & 1.5127 & & 1.5708 & 1.3987 \\
\multicolumn{2}{c}{} & & DGRec & & {0.2964} & 0.3047 & & 2.5203 & {3.8800} & & {3.6386} & {5.5896} & & 2.6931 & 4.8046 & & {8.8520} & 14.2742 & & 1.0708 & 1.0633 & & 0.7719 & 1.7013 & & {1.6061} & 1.4549 \\
\multicolumn{2}{c}{} & & Diversity-focused CPGRec & & \underline{0.3173} & \underline{0.3740} & & \textbf{3.0929} & \textbf{4.7109} & & \textbf{3.8044} & \textbf{6.9557} & & \textbf{3.2404} & \textbf{5.9743} & & \textbf{10.1377} & \textbf{17.6409} & & \textbf{1.2602} & \textbf{1.1937} & & \textbf{0.9054} & \textbf{2.1125} & & \textbf{2.0886} & \textbf{1.6275} \\
\multicolumn{2}{c}{} & & Diversity-focused CPGRec+ & & \textbf{0.3252} & \textbf{0.3845} & & \underline{3.0915} & \underline{4.6929} & & \underline{3.7941} & \underline{6.9465} & & \underline{3.2352} & \underline{5.9678} & & \underline{10.1208} & \underline{17.6072} & & \underline{1.2571} & \underline{1.1921} & & \underline{0.9009} & \underline{2.1083} & & \underline{2.0848} & \underline{1.6254} \\
\midrule
\multicolumn{2}{c}{\multirow{3}{*}{\begin{tabular}[c]{@{}c@{}}Balance-driven\\ Methods\end{tabular}}} & & EXPLORE & & {0.2952} &  {0.3508} & & {2.6180} &  {4.2931} & & {3.3212} &  {5.9811} & & {2.6833} &  {5.2119} & & {8.6225} &  {15.4861} & & {1.0622} &  {1.1423} & & {0.8942} &  {1.4428} & & {1.7854} &  {1.5612} \\
\multicolumn{2}{c}{} & & CPGRec (trade-off framework) & & \underline{0.2963} & \underline{0.3515} & & \textbf{2.8635} & \textbf{4.5731} & & \textbf{3.7032} & \textbf{6.8842} & & \underline{2.6864} & \underline{5.2506} & & \textbf{9.2530} & \textbf{16.7079} & & \textbf{1.0971} & \underline{1.1697} & & \underline{0.9070} & \textbf{1.8828} & & \textbf{1.9418} & \textbf{1.6297} \\
\multicolumn{2}{c}{} & & CPGRec+ (trade-off framework) & & \textbf{0.3047} & \textbf{0.3611} & & \underline{2.7820} & \underline{4.3553} & & \underline{3.6859} & \underline{6.6449} & & \textbf{2.7582} & \textbf{5.2616} & & \underline{9.2261} & \underline{16.2618} & & \underline{1.0835} & \textbf{1.1873} & & \textbf{1.0680} & \underline{1.8666} & & \underline{1.9095} & \underline{1.6171} \\
\bottomrule
\end{tabular}
\end{adjustbox}
\label{table: performance comparison with new metrics: C,E on steam I}
\end{table*}

\begin{table*}[htbp]
\centering
\caption{Diversity-focus performance comparison of different accuracy-, diversity- and balance-driven recommender systems in terms of \textit{Conventional Coverage (CC), Coverage (C)} and \textit{Entropy (E)} on Steam II.}
\begin{adjustbox}{max width = \textwidth}
\begin{tabular}{@{}clcccccccccccccccccccccccccccccccccc@{}}
\toprule
\multicolumn{2}{c}{\multirow{2}{*}{}} & & \multirow{2}{*}{Methods} & & \multicolumn{2}{c}{CC} & & \multicolumn{2}{c}{C(genre)} & & \multicolumn{2}{c}{C(developer)} & & \multicolumn{2}{c}{C(publisher)} & & \multicolumn{2}{c}{C(total)} & & \multicolumn{2}{c}{E(genre)} & & \multicolumn{2}{c}{E(developer)} & & \multicolumn{2}{c}{E(publisher)} \\
\cmidrule(lr){6-7} \cmidrule(lr){9-10} \cmidrule(lr){12-13} \cmidrule(lr){15-16} \cmidrule(lr){18-19} \cmidrule(lr){21-22} \cmidrule(lr){24-25} \cmidrule(lr){27-28}
\multicolumn{2}{c}{} & & & & @5 & @10 & & @5 & @10 & & @5 & @10 & & @5 & @10 & & @5 & @10 & & @5 & @10 & & @5 & @10 & & @5 & @ 10\\
\midrule
\multicolumn{2}{c}{\multirow{7}{*}{\begin{tabular}[c]{@{}c@{}}Accuracy-driven\\ Methods\end{tabular}}} & & LightGCN & & 1.4659 & 1.9124 & & 5.3448 & 6.7381 & & 4.1693 & 8.4357 & & 4.1226 & 7.9835 & & 13.6367 & 23.1573 & & 1.4708 & 1.5936 & & 1.3781 & 2.0019 & & 1.3673 & 1.9382 \\
\multicolumn{2}{c}{} & & SURGE & & 1.4237 & 1.8649 & & 5.2015 & 6.5728 & & 4.0718 & 8.3013 & & 4.0329 & 7.8446 & & 13.3062 & 22.7187 & & 1.4491 & 1.5642 & & 1.3533 & 1.9678 & & 1.3416 & 1.9085 \\
\multicolumn{2}{c}{} & & BIGCF & & 1.4826 & 1.9531 & & 5.4017 & 6.8092 & & 4.2074 & 8.5146 & & 4.1738 & 8.0772 & & 13.7827 & 23.4010 & & 1.4815 & 1.6094 & & 1.3899 & 2.0248 & & 1.3791 & 1.9537 \\
\multicolumn{2}{c}{} & & MVGNN & & 1.4481 & 1.8973 & & 5.2891 & 6.6834 & & 4.1256 & 8.3819 & & 4.0883 & 7.9241 & & 13.5030 & 22.9894 & & 1.4627 & 1.5815 & & 1.3684 & 1.9892 & & 1.3598 & 1.9258 \\
\multicolumn{2}{c}{} & & SCGRec & & 1.4913 & 1.9688 & & 5.4329 & 6.8415 & & 4.2381 & 8.5773 & & 4.2047 & 8.1349 & & 13.8757 & 23.5537 & & 1.4892 & 1.6183 & & 1.3986 & 2.0351 & & 1.3875 & 1.9664 \\
\multicolumn{2}{c}{} & & Accuracy-focused CPGRec & & \underline{1.5714} & \underline{2.0581} & & \textbf{5.8733} & \textbf{7.2498} & & \underline{4.5791} & \underline{9.1206} & & \underline{4.5263} & \underline{8.6141} & & \underline{14.9987} & \underline{24.9845} & & \underline{1.5688} & \underline{1.6993} & & \textbf{1.4839} & \underline{2.1482} & & \underline{1.4728} & \textbf{2.0799} \\
\multicolumn{2}{c}{} & & Accuracy-focused CPGRec+ & & \textbf{1.5752} & \textbf{2.0645} & & \underline{5.8698} & \underline{7.2315} & & \textbf{4.5824} & \textbf{9.1353} & & \textbf{4.5317} & \textbf{8.6256} & & \textbf{15.0211} & \textbf{25.0124} & & \textbf{1.5706} & \textbf{1.7018} & & \underline{1.4801} & \textbf{2.1534} & & \textbf{1.4759} & \underline{2.0763} \\
\midrule
\multicolumn{2}{c}{\multirow{7}{*}{\begin{tabular}[c]{@{}c@{}}Diversity-driven\\ Methods\end{tabular}}} & & MMR & & 1.7351 & 2.2803 & & 5.8691 & 7.3745 & & 4.6527 & 9.1738 & & 4.5019 & 8.7651 & & 15.0237 & 25.3134 & & 1.5642 & 1.7288 & & 1.5026 & 2.1617 & & 1.4664 & 2.1029 \\
\multicolumn{2}{c}{} & & EDUA & & 1.7198 & 2.2581 & & 5.8033 & 7.3016 & & 4.6094 & 9.1025 & & 4.4537 & 8.7039 & & 14.8664 & 25.1080 & & 1.5508 & 1.7134 & & 1.4891 & 2.1462 & & 1.4538 & 2.0887 \\
\multicolumn{2}{c}{} & & DDGraph & & 1.7082 & 2.2436 & & 5.7584 & 7.2583 & & 4.5771 & 9.0496 & & 4.4129 & 8.6572 & & 14.7484 & 24.9651 & & 1.5416 & 1.7027 & & 1.4785 & 2.1331 & & 1.4429 & 2.0754 \\
\multicolumn{2}{c}{} & & DGCN & & 1.7486 & 2.2954 & & 5.9127 & 7.4299 & & 4.6839 & 9.2214 & & 4.5388 & 8.8193 & & 15.1354 & 25.4660 & & 1.5739 & 1.7402 & & 1.5118 & 2.1749 & & 1.4751 & 2.1138 \\
\multicolumn{2}{c}{} & & DGRec & & 1.7591 & 2.3082 & & 5.9438 & 7.4613 & & 4.7015 & 9.2588 & & 4.5631 & 8.8527 & & 15.2084 & 25.5810 & & 1.5813 & 1.7486 & & 1.5194 & 2.1853 & & 1.4829 & 2.1245 \\
\multicolumn{2}{c}{} & & Diversity-focused CPGRec & & \underline{1.8803} & \underline{2.4812} & & \underline{6.1794} & \underline{7.8481} & & \textbf{5.0638} & \underline{9.9682} & & \underline{4.8714} & \underline{9.4753} & & \textbf{16.3015} & \textbf{27.5133} & & \underline{1.6572} & \underline{1.8291} & & \textbf{1.6193} & \underline{2.3024} & & \underline{1.5908} & \underline{2.2491} \\
\multicolumn{2}{c}{} & & Diversity-focused CPGRec+ & & \textbf{1.8851} & \textbf{2.4878} & & \textbf{6.1856} & \textbf{7.8542} & & \underline{5.0571} & \textbf{9.9759} & & \textbf{4.8797} & \textbf{9.4828} & & \underline{16.2897} & \underline{27.4989} & & \textbf{1.6599} & \textbf{1.8324} & & \underline{1.6151} & \textbf{2.3077} & & \textbf{1.5946} & \textbf{2.2562} \\
\midrule
\multicolumn{2}{c}{\multirow{3}{*}{\begin{tabular}[c]{@{}c@{}}Balance-driven\\ Methods\end{tabular}}} & & EXPLORE & & 1.6904 & 2.2158 & & 5.6923 & 7.1504 & & 4.5018 & 8.8421 & & 4.3791 & 8.4632 & & 14.5732 & 24.4557 & & 1.5187 & 1.6733 & & 1.4562 & 2.0914 & & 1.4206 & 2.0319 \\
\multicolumn{2}{c}{} & & CPGRec (trade-off framework) & & \textbf{1.7749} & \underline{2.3216} & & \textbf{5.9994} & \underline{7.5189} & & \underline{4.7381} & \textbf{9.3046} & & \underline{4.5892} & \underline{8.8805} & & \underline{15.3108} & \underline{25.6943} & & \underline{1.5921} & \underline{1.7549} & & \textbf{1.5298} & \underline{2.1984} & & \textbf{1.4941} & \underline{2.1388} \\
\multicolumn{2}{c}{} & & CPGRec+ (trade-off framework) & & \underline{1.7725} & \textbf{2.3251} & & \underline{5.9863} & \textbf{7.5241} & & \textbf{4.7458} & \underline{9.2993} & & \textbf{4.5937} & \textbf{8.8898} & & \textbf{15.3259} & \textbf{25.7132} & & \textbf{1.5948} & \textbf{1.7573} & & \underline{1.5264} & \textbf{2.2031} & & \underline{1.4916} & \textbf{2.1425} \\
\bottomrule
\end{tabular}
\end{adjustbox}
\label{table: performance comparison with new metrics: C,E on steam II}
\end{table*}

\begin{table*}[htbp]
\centering
\caption{Diversity-focus performance comparison of different accuracy-, diversity- and balance-driven recommender systems in terms of \textit{Tail Coverage} and \textit{Tail} on Steam I.}
\begin{adjustbox}{max width = 0.5\textwidth}
\begin{tabular}{@{}l l cccc@{}}
\toprule
\multicolumn{2}{c}{\multirow{2}{*}{Method}} & \multicolumn{2}{c}{Tail Coverage} & \multicolumn{2}{c}{Tail} \\
\cmidrule(lr){3-4} \cmidrule(lr){5-6}
\multicolumn{2}{c}{} & @5 & @10 & @5 & @10 \\
\midrule
\multirow{7}{*}{\begin{tabular}[c]{@{}l@{}}Accuracy-driven\\ \hspace{0.5em}Methods\end{tabular}}
& LightGCN & 0.1364 & 0.1626 & 0.0251 & 0.0036 \\
& SURGE & 0.1383 & 0.1664 & 0.0248 & 0.0035 \\
& BIGCF & 0.1421 & 0.1738 & 0.0263 & 0.0037 \\
& MVGNN & 0.1402 & 0.1813 & 0.0257 & 0.0037 \\
& SCGRec & 0.1346 & 0.1607 & 0.0250 & 0.0036 \\
& Accuracy-focused CPGRec & \underline{0.1458} & \underline{0.2075} & \underline{0.0279} & \underline{0.0037} \\
& Accuracy-focused CPGRec+ & \textbf{0.1495} & \textbf{0.2280} & \textbf{0.0280} & \textbf{0.0038} \\
\midrule
\multirow{7}{*}{\begin{tabular}[c]{@{}l@{}}Diversity-driven\\ \hspace{0.5em}Methods\end{tabular}}
& MMR & 0.1308 & 0.1607 & 0.0251 & 0.0036 \\
& EDUA & 0.1346 & 0.1645 & 0.0250 & 0.0036 \\
& DDGraph & 0.1364 & 0.1645 & 0.0250 & 0.0036 \\
& DGCN & 0.1421 & 0.1645 & 0.0278 & 0.0037 \\
& DGRec & 0.1402 & 0.2112 & 0.0283 & 0.0035 \\
& Diversity-focused CPGRec & \underline{0.1794} & \underline{0.2561} & \underline{0.0362} & \underline{0.0051} \\
& Diversity-focused CPGRec+ & \textbf{0.1869} & \textbf{0.2654} & \textbf{0.0375} & \textbf{0.0057} \\
\midrule
\multirow{3}{*}{\begin{tabular}[c]{@{}l@{}}Balance-driven\\ \hspace{0.5em}Methods\end{tabular}}
& EXPLORE & 0.1477 & 0.2243 & 0.0277 & 0.0036 \\
& CPGRec (trade-off framework) & \underline{0.1533} & \underline{0.2336} & \underline{0.0284} & \underline{0.0038} \\
& CPGRec+ (trade-off framework) & \textbf{0.1682} & \textbf{0.2617} & \textbf{0.0308} & \textbf{0.0047} \\
\bottomrule
\end{tabular}
\end{adjustbox}
\label{table: performance comparison tail metrics on steam I}
\end{table*}

\begin{table*}[htbp]
\centering
\caption{Diversity-focus performance comparison of different accuracy-, diversity- and balance-driven recommender systems in terms of \textit{Tail Coverage} and \textit{Tail} on Steam II.}
\begin{adjustbox}{max width = 0.5\textwidth}
\begin{tabular}{@{}l l cccc@{}}
\toprule
\multicolumn{2}{c}{\multirow{2}{*}{Method}} & \multicolumn{2}{c}{Tail Coverage} & \multicolumn{2}{c}{Tail} \\
\cmidrule(lr){3-4} \cmidrule(lr){5-6}
\multicolumn{2}{c}{} & @5 & @10 & @5 & @10 \\
\midrule
\multirow{7}{*}{\begin{tabular}[c]{@{}l@{}}Accuracy-driven\\ \hspace{0.5em}Methods\end{tabular}}
& LightGCN & 0.3281 & 0.4214 & 0.1495 & 0.3188 \\
& SURGE & 0.3315 & 0.4287 & 0.1489 & 0.3172 \\
& BIGCF & 0.3402 & 0.4416 & 0.1517 & 0.3241 \\
& MVGNN & 0.3378 & 0.4503 & 0.1506 & 0.3253 \\
& SCGRec & 0.3259 & 0.4192 & 0.1491 & 0.3194 \\
& Accuracy-focused CPGRec & \underline{0.3501} & \underline{0.4625} & \underline{0.1583} & \underline{0.3419} \\
& Accuracy-focused CPGRec+ & \textbf{0.3507} & \textbf{0.4710} & \textbf{0.1610} & \textbf{0.3477} \\
\midrule
\multirow{7}{*}{\begin{tabular}[c]{@{}l@{}}Diversity-driven\\\hspace{0.5em}Methods\end{tabular}}
& MMR & 0.3305 & 0.4311 & 0.1513 & 0.3204 \\
& EDUA & 0.3348 & 0.4382 & 0.1511 & 0.3209 \\
& DDGraph & 0.3371 & 0.4390 & 0.1512 & 0.3211 \\
& DGCN & 0.3426 & 0.4401 & 0.1593 & 0.3284 \\
& DGRec & 0.3411 & 0.4658 & 0.1612 & 0.3301 \\
& Diversity-focused CPGRec & \underline{0.3586} & \underline{0.4791} & \underline{0.1654} & \underline{0.3562} \\
& Diversity-focused CPGRec+ & \textbf{0.3607} & \textbf{0.4860} & \textbf{0.1703} & \textbf{0.3647} \\
\midrule
\multirow{3}{*}{\begin{tabular}[c]{@{}l@{}}Balance-driven\\\hspace{0.5em}Methods\end{tabular}}
& EXPLORE & 0.3453 & 0.4618 & 0.1589 & 0.3345 \\
& CPGRec (trade-off framework) & \underline{0.3512} & \underline{0.4693} & \underline{0.1615} & \underline{0.3497} \\
& CPGRec+ (trade-off framework) & \textbf{0.3514} & \textbf{0.4766} & \textbf{0.1646} & \textbf{0.3538} \\
\bottomrule
\end{tabular}
\end{adjustbox}
\label{table: performance comparison tail metrics on steam II}
\end{table*}

An exhaustive comparison of CPGRec+ against various baseline models is detailed in Tables \ref{table: performance comparison with new metric: accu and cc on steam I}, \ref{table: performance comparison with new metric: accu and cc on steam II}, \ref{table: performance comparison with new metrics: C,E on steam I}, \ref{table: performance comparison with new metrics: C,E on steam II}, \ref{table: performance comparison tail metrics on steam I}, and \ref{table: performance comparison tail metrics on steam II}. The superior and runner-up outcomes are distinctively highlighted in bold and underlined text, respectively. Drawing from these experimental findings, we can distill the following insights:

\begin{itemize}
\item \textbf{Superiority of the Proposed Method w.r.t Accuracy or Diversity.} (1) CPGRec+, in its accuracy-focused form, outperforms all other state-of-the-art accuracy-centric methods across all metrics. (2) Similarly, the diversity-focused CPGRec+ surpasses all diversity-centric approaches except for CPGRec, which serves as a variant of CPGRec+ by removing \textcolor{black}{the two newly proposed modules (both PER and PRG}), on every metric. \textcolor{black}{Moreover, when compared with CPGRec, CPGRec+ consistently outperforms CPGRec among long-tail-centric metrics like \textit{Conventional Coverage}, \textit{Tail Coverage}, and \textit{Tail} due to its personal-preference-centeredness, which mitigates the dominance of popular games and fosters the long-tail game recommendations though slightly restricting the categorical diversity indicated by \textit{Coverage} and \textit{Entropy}.} In essence, these results underscore CPGRec+'s superior performance in both accuracy and diversity.

\item \textbf{Superiority of the Proposed Method w.r.t Capability of Flexible Adjusting.} CPGRec+ along with CPGRec have outperformed the SOTA in the domains of accuracy, diversity, and balance, respectively. This demonstrates that the supposed dichotomy between accuracy and diversity is not inevitable. A well-calibrated recommender system like CPGRec+ can effectively harmonize these two aspects, precisely capturing user preferences while delivering a diverse selection that enriches the player experience.

\item \textbf{Double-Edged Nature of Graph Convolutional Smoothness.} (1) On one hand, effectively leveraging smoothness encodes category information into the representations of video games through graph structural transformations, thereby enhancing accuracy. This strategy is evident in SCGRec and both the SGC and CNA modules of CPGRec+; (2) on the other hand, the smoothness resulting from stacking GCN layers reduces the distinctiveness of representations between neighboring games, which in turn affects the improvement of accuracy, as seen in diversity-driven baselines like DGRec. The stark contrast between the aforementioned two aspects is not surprising, as the latter's design motivation is diversity rather than accuracy.

\end{itemize}
On top of that, \textcolor{black}{by comparing CPGRec+ and CPGRec}, it is worth noting that \textcolor{black}{our newly introduced modules (PER and PRG)} not only show significant empowerment of accuracy, but also improve remarkedly the diversity illustrated by \textit{Conventional Coverage} and the two long-tail-centric metrics \textit{Tail Coverage} and \textit{Tail} . Detailed analysis is deferred to the following ablation study presented in Section \ref{sec: ablation study}.

\subsection{Ablation Study}\label{sec: ablation study}
Within this segment, we conduct an ablation analysis on CPGRec+ on Steam I by methodically removing each of its five modules. For the SGC module, we introduce two variants: one involves the complete removal of the SGC module, while the other substitutes strict connections with raw ones. For the PER module, we nullify its effect by assigning a preference-informed weight of 0 to each edge within the player-game graph. Our observations from the experimental results, presented in Table \ref{table: ablation}, are as follows:

\begin{itemize}
    \item \textcolor{black}{\textbf{Effectiveness of Existing Four Modules.}}
    The observations from removing each of the existing four modules (SGC, CNA, PENR, and NSR), as presented in our conference work \cite{cpgrec}, are consistent with the results from that publication, so a redundant analysis is omitted here. However, it is important to note that these results not only validate the efficacy of the existing four modules but also confirm the compatibility of the newly introduced PER and PRG modules with them.

    \item \textbf{Effectiveness of PER Module w.r.t Accuracy.} The removal of the PER module significantly degraded CPGRec+'s performance across all accuracy metrics, underscoring PER's pivotal role in tailoring recommendations to player interests. This indicates that PER effectively provides a granular modeling of a player's preferences by contrasting the personal interest in each historical record with the global interests of the public.

    \item \textcolor{black}{
    \textbf{Effectiveness of PRG Module w.r.t Accuracy.}
    Similarly, removing PRG results in a significant decline in the accuracy of CPGRec+. This is expected, as the rich knowledge embedded in LLMs enables the generation of more comprehensive game descriptions beyond the dataset. Additionally, the reasoning capabilities of LLMs not only identify deep commonalities among games from multiple perspectives but also infer a player's personal preferences, effectively constructing a more informative player profile.
    }

    \item \textcolor{black}{\textbf{Both PER and PRG Foster Balance-oriented Recommendations.}}
     Although the removals of the PER and PRG both result in a modest enhancement in Coverage@10 as well as a significant retrogression across the accuracy metrics relative to CPGRec+, which aligns with the anticipated trade-off between accuracy and diversity observed in numerous empirical studies \cite{kunaver2017diversity,shi2023relieving,javari2015probabilistic,peng2024reconciling, 10.1145/3638352}; \textcolor{black}{moreover, it is worth noting that a marked reduction is evident in both the other two diversity metrics, \textit{i.e.,} Conventional Coverage and Tail. This finding highlights the critical role of comparing individual player preferences with public interests, which are employed in both PER and PRG: in terms of accuracy, the model gain deeper understanding of the players' inherent interests by reasoning and comparing, which benefits more accurate recommendations; in terms of diversity, the model's ability to reason about players' inherent preferences allows it to better identify suitable long-tail games to be recommended, thereby enhancing long-tail recommendations.}

        \item \textcolor{black}{\textbf{PENR harmonizes different perspectives of diversity.} Coverage (C), Conventional Coverage (CC), and Tail Coverage are three diversity metrics from different perspectives: C measures the breadth of categories covered by recommended games, while CC and Tail Coverage measure the number of games and long-tail games covered by the recommended lists, respectively. The essential incorporation of the Popularity-guided Edges and Nodes Reweighting (PENR) component effectively fosters a balance in CPGRec+'s performance across these three perspectives on diversity, emphasizing the crucial objective of \textbf{long-tail recommendation} in this study. PENR's mechanism of reweighting the player-game graph amplifies the influence of long-tail games, which not only boosts Tail Coverage directly but also brings the cascading effect of implicitly diversifying the categories of top 5 recommended games measured by Coverage, while keeping the Conventional Coverage just a hair behind the suboptimal levels.}
\end{itemize}

\begin{table*}[htbp] 
 \centering 
 \caption{Ablation study. We show CPGRec+'s performance when removing each of the modules.} 
 \begin{adjustbox}{max width = \textwidth} 
 \begin{tabular}{@{}lcccccccccccc@{}} 
 \toprule 
 \multicolumn{1}{c}{Method} & Recall@5 & Recall@10 & Hit@5 & Hit@10 & Precision@5 & Precision@10 & C(total)@5 & C(total)@10 & CC@5 & CC@10 & Tail Coverage@5 & Tail Coverage@10 \\ \midrule 
 \textbf{CPGRec+} (\textbf{trade-off} framework) & \underline{0.5347} & \underline{0.6192} & \underline{0.6590} & \underline{0.7473} & \underline{0.1538} & \underline{0.0967} & \underline{9.2261} & 16.2618  & 0.3047 & 0.3611 & \underline{0.1682} & \textbf{0.2617} \\ 
 \midrule 
 w/o SGC (1st accuracy-driven module) & 0.5284 & 0.6114 & 0.6506 & 0.7404 & 0.1508 & 0.0947 & \textbf{9.7154} & \textbf{17.3755} & \underline{0.3076} & \underline{0.3654} & \textbf{0.1720} & \underline{0.2542} \\ 
 CPGRec+ (SGC with raw connections) & 0.5275 & 0.6114 & 0.6502 & 0.7407 & 0.1499 & 0.0941 & 8.4256 & 15.6349 & 0.2959 & 0.3490 & 0.1664 & 0.2449 \\ 
 w/o \textbf{PER} (2nd accuracy-driven module) & 0.5277 & 0.6119 & 0.6498 & 0.7406 & 0.1501 & 0.0946 & 9.2041 & \underline{16.7189} & 0.2924 & 0.3468 & 0.1607 & 0.2393 \\ 
 w/o \textbf{PRG} (3rd accuracy-driven module) & 0.5252 & 0.6101 & 0.6481 & 0.7402 & 0.1498 & 0.0944 & 9.0745 & 16.5235 & 0.2895 & 0.3402 & 0.1589 & 0.2374 \\ 
 \midrule 
 w/o CNA (1st diversity-driven module)  & 0.5311 & 0.6106 & 0.6545 & 0.7393 & 0.1518 & 0.0947 & 8.3133 & 14.3171 & 0.2799 & 0.3318 & 0.1551 & 0.2355 \\ 
 w/o PENR (2nd diversity-driven module)  & \textbf{0.5794} & \textbf{0.6376} & \textbf{0.7055} & \textbf{0.7629} & \textbf{0.1630} & \textbf{0.0989} & 8.7021 & 16.3384 & \textbf{0.3148} & \textbf{0.3765} & 0.1589 & 0.2393 \\ 
 \midrule 
 w/o NSR (comprehensive module)& 0.5131 & 0.6018 & 0.6333 & 0.7330 & 0.1491 & 0.0924 & 7.8537 & 13.2116 & 0.2832 & 0.3348 & 0.1570 & 0.2374 \\ \bottomrule 
 \end{tabular} 
 \end{adjustbox} 
 \label{table: ablation}  
 \end{table*}

\begin{figure}[htbp]
    \begin{minipage}{0.24\textwidth}
    \centering
    \includegraphics[width=\textwidth]{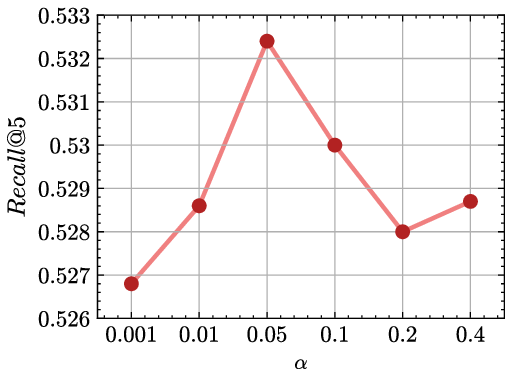}
    \end{minipage} %
    \begin{minipage}{0.24\textwidth}
    \centering \hspace{-0mm}
    \includegraphics[width=\textwidth]{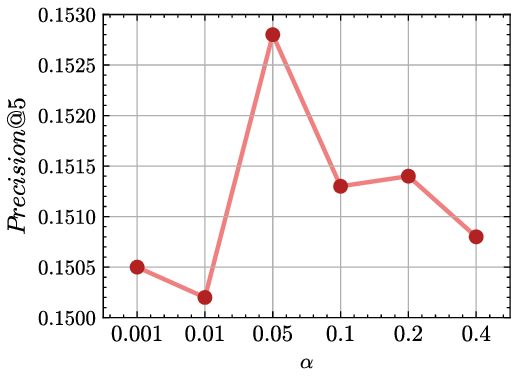}
    \end{minipage} %
    \begin{minipage}{0.24\textwidth}
    \centering
    \includegraphics[width=\textwidth]{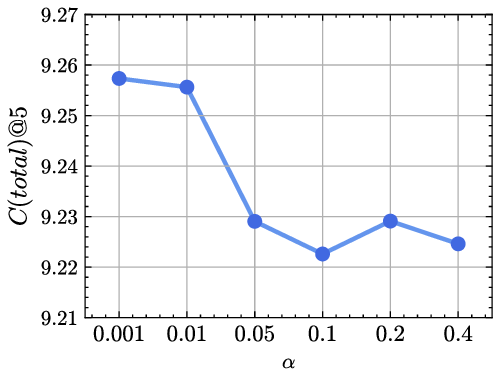}
    \end{minipage} %
    \begin{minipage}{0.243\textwidth}
    \centering
    \includegraphics[width=\textwidth]{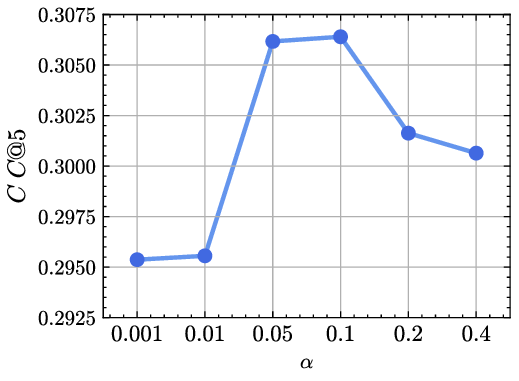}
    \end{minipage} %
\caption{Parameter sensitivity on $\alpha$, which is the significance level applied in PER module.}\vspace{-5mm}
\label{fig:alpha sens}
\end{figure}

\subsection{Parameter Sensitivity}

\textcolor{black}{In this section, we analyze the impact of the hyper-parameter $\alpha$ in the PER module on Steam I. The parameter $\alpha$ represents the significance level and determines the operational scope of PER by setting the threshold $Q_{\alpha}$ for the Fisher distribution-based sign decision. Specifically, a smaller $\alpha$ indicates stricter conditions for applying preference-informed edge reweighting, requiring the player to demonstrate a more significant deviation from the global preference of general players. Conversely, a larger $\alpha$ implies more lenient conditions.}

To evaluate the effectiveness of $\alpha$, we conducted experiments by varying its value within the range $[0.001, 0.01, 0.05, 0.1, 0.2, 0.4]$. For accuracy assessment, we utilized \textit{Recall@5} and \textit{Precision@5}; for diversity evaluation, we employed \textit{Coverage@5} and \textit{Conventional Coverage@5}. The experimental results, presented in Fig. \ref{fig:alpha sens}, demonstrate that $\alpha$ significantly influences both accuracy and diversity, with optimal performance observed at moderate values of $\alpha$.

\textcolor{black}{
For \textbf{accuracy}, the analysis reveals two distinct trends:
\begin{enumerate}
    \item When $\alpha$ is small (specifically, $\alpha \in \{0.001, 0.01, 0.05\}$), both \textit{Recall} and \textit{Precision} increase as $\alpha$ rises. Notably, a fairly small $\alpha$ imposes strict conditions for using the PER module, similar to not applying it, while increasing $\alpha$ moderately broadens the scope for preference-informed weights. This underscores the critical role of $\alpha$ in enhancing recommendation accuracy.
    \item However, as $\alpha$ grows larger ($\alpha \in \{0.05, 0.1, 0.2, 0.4\}$), both \textit{Recall} and \textit{Precision} decline. A larger $\alpha$ makes PER’s conditions too lenient, including less reliable preferences, which adds noise and reduces accuracy.
\end{enumerate}
For \textbf{diversity}, the metrics \textit{Coverage} and \textit{Conventional Coverage} behave differently as $\alpha$ changes:
\begin{enumerate}
    \item As $\alpha$ increases, \textit{Coverage} first decreases, then rises. This differs from accuracy trends, showing a trade-off: wider-ranging recommendations can improve diversity but may lower accuracy.
    \item Meanwhile, \textit{Conventional Coverage} rises initially with $\alpha$, then slightly drops, similar to \textit{Recall} and \textit{Precision}. This occurs because the PER module uses dwelling time and game ratings to better model players and games, making recommendations more unique and globally diverse.
\end{enumerate}
}

\begin{figure}[t]
\begin{minipage}{0.618\textwidth}
\centering
\includegraphics[width=\textwidth]{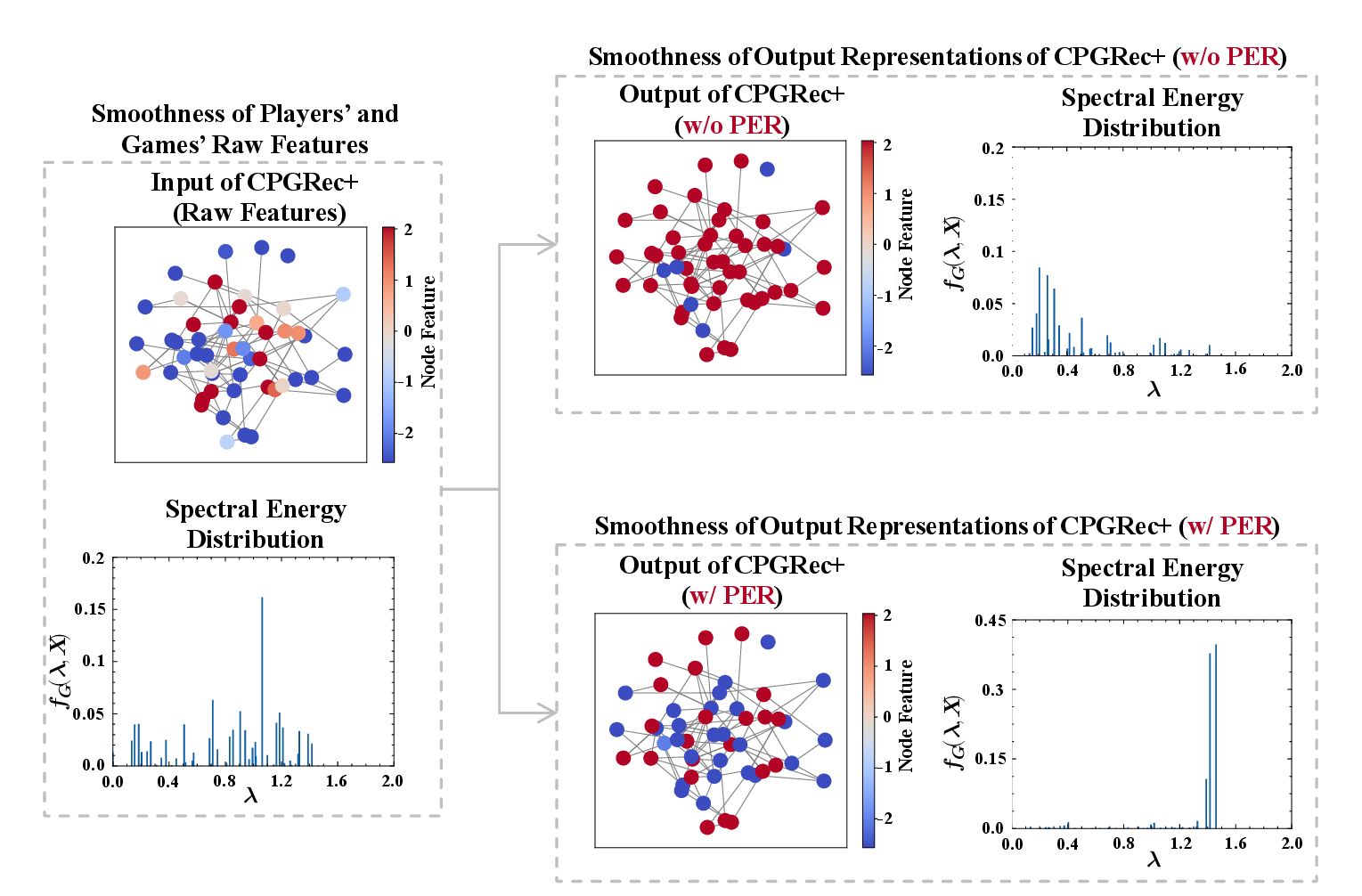}
\caption{Spectral energy distribution of node representations on the player-game game. The representations on the left figure is the raw feature of players and games, and those on the right are respectively processed by CPGRec+ w/o PER and w/ PER.}\vspace{-5mm}
\label{fig:smoothness}
\end{minipage} 
\end{figure}

\subsection{Intuitive Understanding of the Functionality of PER} \label{sec: intuitive understanding of per}
In this section, we provide a detailed explanation to enhance understanding of the PER module's role within CPGRec+. We use a subgraph from the player-game bipartite graph as an illustrative example. This subgraph consists of 50 nodes representing players or games, with edges denoting historical interactions. Each node is assigned a 1-dimensional simulated feature. To demonstrate CPGRec+'s functionality, we compare it to a variant without the PER module, focusing on how each processes input features. Fig. \ref{fig:smoothness} presents the subgraph details, including spectral energy distributions as a quantitative supplement. The spectral energy distribution is defined as:
\begin{align}
    f_G(\Lambda,X) &= \big[f_G(\lambda_1,X),f_G(\lambda_2,X),\cdots,f_G(\lambda_{\mathcal{N}},X)\big]\\    
    &=\Big[\frac{\widetilde{x}_1^T\widetilde{x}_1}{\sum_{i=1}^{\mathcal{N}}\widetilde{x}_i^T\widetilde{x}_i},
    \frac{\widetilde{x}_2^T\widetilde{x}_2}{\sum_{i=1}^{\mathcal{N}}\widetilde{x}_i^T\widetilde{x}_i},
    \cdots,
    \frac{\widetilde{x}_{\mathcal{N}}^T\widetilde{x}_{\mathcal{N}}}{\sum_{i=1}^{\mathcal{N}}\widetilde{x}_i^T\widetilde{x}_i}\Big],
\end{align}
where $\widetilde{x}_i$ is the dot product of the $i$-th normalized eigenvector $u_i$ defined as Equation \ref{eq: eigenvector} and the nodes' feature $X\in \mathcal{R}^{\mathcal{N}}$. Smaller $\lambda_i$ values correspond to low-frequency components, and vice versa. The smoothness of $X$ is characterized by the spectral energy distribution's emphasis on low-frequency parts: higher $f_G(\lambda_i,X)$ values for smaller $\lambda_i$ indicate greater smoothness, while higher weights in mid to high frequencies suggest less smoothness.

The left figure in Fig. \ref{fig:smoothness} illustrates the raw feature smoothness, showing distinct node features with non-uniform colors. The spectral energy distribution suggests a significant distribution across low and mid frequencies.

The right figures in Fig. \ref{fig:smoothness} compare representations processed by CPGRec+ without and with the PER module, with observations as follows:

(1) The first row shows that without the PER module, representations exhibit over-smoothness, as indicated by uniform node colors (\textit{e.g.}, red) and concentrated low-frequency spectral energy. This aligns with Section \ref{section: disparity}'s analysis, where neglecting interaction disparities leads to over-smoothness due to graph convolutional layers' inherent smoothness.

(2) Conversely, the second row demonstrates that incorporating the PER module helps CPGRec+ avoid over-smoothness, evident from varied node colors. The spectral energy distribution retains high-frequency components, indicating that CPGRec+ captures preference-informed personal interest and disinterest, extending beyond a low-pass filter to address personalized player needs.

\textcolor{black}{
\subsection{Analysis of Fusion Methods} \label{sec: fusion analysis}
As detailed in the introduction to the component PRG in Section \ref{sec: new module: PRG}, MLPs are leveraged to fuse the embedded representation of players and games generated by LLMs and the corresponding original representations to generate more informative final representations $e_u$ and $e_i$, respectively. In this section, we meticulously compare the performance of the proposed model when combined with different fusion methods, including linear and gated fusion, to provide more complete analysis and insights. The experimental results from Steam I are presented in Table \ref{table: fusion analysis}.
\\
\\
As seen in Table \ref{table: fusion analysis}, when combined with MLP, CPGRec+ significantly outperforms the other two variants with linear and gated fusion methods in terms of both accuracy and diversity across most cases, which justifies the choice of MLP as the default fusion method. Besides, based on the following two observations, the experimental results can be attributed to the ability of different fusion methods to capture deep player preference information: (1) the MLP model has more learnable parameters and the ability to capture higher-order feature interactions than linear and gated fusion methods to fit the complex patterns of player-game interaction; (2) gated fusion significantly outperforms the linear fusion since it can not only adaptively adjust the weights of the original representation and the representation generated by LLM, but also introduces a nonlinear control mechanism through the gated network, which has stronger expressive power than linear fusion.
}

\begin{table*}[htbp]
\centering
\caption{CPGRec+'s performance when it is combined with different fusion methods.}
\begin{adjustbox}{max width = \textwidth}
\begin{tabular}{@{}lcccccccccc@{}}
\toprule
\multicolumn{1}{c}{Fusion Method} & Recall@5 & Recall@10 & Hit@5 & Hit@10 & Precision@5 & Precision@10 & C(total)@5 & C(total)@10 & CC@5 & CC@10 \\ \midrule
MLP (default) & \textbf{0.5347} & \textbf{0.6192} & \textbf{0.6590} & \textbf{0.7473} & \textbf{0.1538} & \textbf{0.0967} & \textbf{9.2261} & \underline{16.2618} & \textbf{0.3047} & \textbf{0.3611} \\
\midrule
Linear & 0.5170 & 0.6023 & 0.6382 & 0.7316 & 0.1472 & 0.0928 & 8.8078 & \textbf{16.2777} & 0.2830 & 0.3372 \\
Gated & \underline{0.5266} & \underline{0.6080} & \underline{0.6467} & \underline{0.7361} & \underline{0.1493} & \underline{0.0936} & \underline{8.9275} & 15.9849 & \underline{0.2935} & \underline{0.3540} \\
\bottomrule
\end{tabular}
\end{adjustbox}
\label{table: fusion analysis}
\end{table*}

\begin{table*}[htbp]
\centering
\caption{Historical interactions of the player $i$. Each game's ID is marked after its name. A game may have multiple IDs representing different versions, \textit{e.g.}, the official version and a free trial or promotional version.}
    \begin{adjustbox}{max width=\textwidth}
            \begin{tabular}{@{}ll@{}}
            \toprule
            Division & Historical Games \\ \midrule
            Train Set & Counter-Strike: Source (240), Day of Defeat: Source (300), Call of Duty®: Modern Warfare® 3 (42680, 42690)\\
            Valid Set & Half-Life 2: Lost Coast (340) \\
            Test Set & Half-Life 2: Deathmatch (320)\\ \bottomrule
            \end{tabular}            
    \end{adjustbox}\label{tab: case study personal records}
\end{table*}

\begin{table*}[htbp]
\centering
\caption{Basic information of the player \textit{i}'s historical games within the train set. Only the information of games within train set is shown since they are the only accessible information to foster CPGRec+’s deeper understanding of player \textit{i}’s preferences.}
    \begin{adjustbox}{max width=\textwidth}
        \begin{tabular}{@{}ccccc@{}}
    \toprule
    Game ID & 240 & 300 & 42680 & 42690 \\ \midrule
    Average Rating & 88 & 80 & 74 & 74 \\
    Dwelling Time & 42 & 91,977 & 124 & 158 \\
    \midrule
    \multicolumn{1}{l}{Mapped Rating} & \multicolumn{1}{l}{1.0232} & \multicolumn{1}{l}{0.0102} & \multicolumn{1}{l}{-0.7678} & \multicolumn{1}{l}{-0.7678} \\
    \multicolumn{1}{l}{Mapped Time} & \multicolumn{1}{l}{-0.8195} & \multicolumn{1}{l}{3.7610} & \multicolumn{1}{l}{-0.4717} & \multicolumn{1}{l}{-0.3849} \\
    \midrule
    Fisher Statistic & 0.6415 & 136,960 & 0.3775 & 0.2514 \\ \bottomrule
    \end{tabular}
    \end{adjustbox}\label{tab: case study game info}
\end{table*}

\begin{table*}[htbp]
\centering
\caption{Experimental results of Top 10 Recommendations for the player \textit{i}. The games within the list are organized in descending order of predicted probability. The target game is highlighted in bold.}
    \begin{adjustbox}{max width=\textwidth}
        \begin{tabular}{@{}cc@{}}
        \toprule
        Setting & Top 10 Recommendations \\ \midrule
        (a) CPGRec & $[42710, 42700, 730, 10090, 340, 202990, 55100, 57900,58200,23500]$ \\
        (b) CPGRec w/ PER & $[570, 20, 630, \textbf{320}, 40, 60, 730, 1520, 440, 202990 ]$ \\
        (c) CPGRec w/ PRG & $[340, 42700, \textbf{320}, 42710, 730,  7940, 12210,380, 630, 60]$ \\
        \multicolumn{1}{l}{(d) CPGRec+ (CPGRec w/ PER, PRG)} & $[\textbf{320}, 340, 440, 20, 630, 30, 570, 1520, 60,42700]$ \\ \bottomrule
        \end{tabular}
    \end{adjustbox}\label{tab: case study experiments}
\end{table*}

\textcolor{black}{
\subsection{Case Study on the Interplay between PER and PRG}
In this section, a detailed case study is provided to demonstrate and analyze the interplay between PER and PRG, the two components newly proposed in this study. Specifically, we compare and analyze the recommendation results of the proposed framework for player \textit{i} (Steam ID = 76561198037067087) on Steam I under four different settings: (a) CPGRec (i.e., CPGRec+ w/o PER and PER), (b) CPGRec w/ PER, (c) CPGRec w/ PRG, (d) CPGRec+ (i.e., CPGRec w/ PER and PRG). The historical games of player \textit{i}, the basic information of these historical games, and the experimental results are shown in Tables \ref{tab: case study personal records}, \ref{tab: case study game info}, and \ref{tab: case study experiments}, respectively.
\\
Based on the experimental results shown in Table \ref{tab: case study experiments}, we have the following key observations for each setting:
\begin{itemize}
    \item \textbf{Regarding setting (a):} two key observations are made. First, the target item (320) is absent from the recommendation list, reflecting the model's limited capacity for providing accurate recommendations. Furthermore, the model ranks Call of Duty®: Black Ops (42710, 42700), Counter-Strike: Global Offensive (730), and Call of Duty: World at War (10090) at the top of the list. This occurs despite the demonstrably limited interest player \textit{i} has shown in other versions of these games (240, 42680, 42690), as evidenced by the dwelling time in Table \ref{tab: case study game info}. This illustrates that in the absence of PER, CPGRec itself lacks the ability to discern from historical interactions those that are more indicative of a player's personalized interests and to learn from them preferentially.
    \item \textbf{Regarding setting (b):} upon incorporating the PER component, the model acquires the ability to discern the disparity in importance among historical interactions, from which it benefits significantly. On the one hand, the model assigns greater attention to the interaction between player \textit{i} and Day of Defeat: Source (300), as its corresponding Fisher statistic is statistically significant, leading to the precise recommendation of the target game (320). On the other hand, the model discerns that the player's historical interactions with the other three games (240, 42680, 42690) are not significantly indicative of its interests, notwithstanding their presence in the historical data. Consequently, while alternative versions of these games (e.g., Counter-Strike: Global Offensive, 730; Call of Duty®: Black Ops II, 202990) are included in the recommendation list, they are assigned considerably lower ranks. Nevertheless, the model's inability to place the target game (320) at the forefront of the recommendation list reveals a deficiency in its predictive precision, which is potentially due to an insufficient grasp of the inherent information pertaining to the games and players. This specific challenge prompted us to incorporate PRG.
    \item \textbf{Regarding setting (c):} upon incorporating PRG, the model gains enhanced insights into all games and players, encoded in their respective representations. Leveraging this, the model ranks the target game (320) and a related version (340) in the top three positions. A persistent issue remains, however, as the model also assigns high ranks to versions of games (42700, 42710, 730) that held no significant interest for player \textit{i} in the historical data (240, 42680, 42690). This suggests that the model is still unable to distinguish the relative importance of past interactions and remains susceptible to the over-smoothing problem inherent in bipartite graphs. It was this challenge that inspired the introduction of PER.
    \item \textbf{Regarding setting (d):} under the interplay of PER and PRG, the model selectively focuses on the more critical historical interactions, leveraging its profound understanding of the games and player preferences to provide ideal recommendations. Specifically, the target game (320) and its variant (340) are ranked in the top two positions, while a different version (42700) of games (42680, 42690) that proved to be less appealing to player \textit{i} is ranked last. In comparison to the player's highly-favored Day of Defeat: Source (300), these recommended games not only feature a diverse range of settings and styles, including cartoon (20, 440), battlefield (320, 1520, 42700), space (630), and epic fantasy (570), but also maintain a consistent core gameplay. They all necessitate that players fulfill distinct roles within a class-based system, devise and execute sophisticated tactics, and engage in close coordination and teamwork.
\end{itemize}
In summary, the PER and PRG interplay enables CPGRec+ to fully grasp the player's preferences and provide accurate and rich recommendations. PER, through improved graph structure, enables CPGRec+ to focus more on historical interactions that significantly reflect individual interests. However, CPGRec+ achieves suboptimal performance without the informative representations provided by the PRG as input. Conversely, CPGRec+ without the PRG inevitably suffers from the dilemma of over-focusing on noisy historical interactions, also at the expense of reduced accuracy.
}

\section{Conclusion}

To address the accuracy-diversity trade-off in video game recommendation, we identify two key limitations in GCN-based models: oversmoothing, which homogenizes recommendations, and the neglect of interaction-level disparities, which weakens personalization. This motivates enhancements to our prior framework, CPGRec, resulting in CPGRec+.


Motivated by these findings, we proposed \textbf{CPGRec+}, which introduces two novel modules to address interaction disparities and balance competing objectives. The \textbf{Preference-informed Edge Reweighting (PER)} module leverages dwell time and ratings—transformed via Box-Cox and Fisher distribution linkages—to differentiate player interests and disinterests, mitigating over-smoothing by reweighting edges with information-theoretic confidence scores. Complementing this, the \textbf{Preference-informed Representation Generation (PRG)} module harnesses LLMs to synthesize game descriptions enriched by global ratings and infer personalized player preferences through comparative reasoning to generate player descriptions, refining embeddings with semantic insights. 
Experiments on real-world data demonstrated CPGRec+’s superiority: it achieves significantly better accuracy than state-of-the-art accuracy-focused models while maintaining diversity metrics comparable to specialized baselines. Ablation studies underscore the necessity of PER and PRG, with their removal causing significant accuracy drops and reduced global coverage.


Future work will explore dynamic graph updates to handle evolving player preferences and cold-start scenarios. Furthermore, as player interests shift over time, ensuring the reliability of LLM-based reasoning becomes critical. Integrating robust uncertainty quantification mechanisms for self-evolving LLMs~\cite{xfzhou} could significantly enhance the trustworthiness of generated player profiles in such dynamic environments. Additionally, integrating LLMs for cross-modal game understanding could further bridge semantic gaps, advancing toward truly adaptive recommendation ecosystems.

To extend our study of representation fusion, future work could test whether low-rank tensor regularization, as used in multi-view clustering~\cite{PRTL}, improves the fusion of graph and text representations. Incomplete multi-view clustering~\cite{HSIMTC} also suggests testing how well this fusion works when one of those representations is unavailable.


\begin{acks}

This work was partially supported by 
the National Natural Science Foundation of China (Project No. 62202122, 62073272, and 62572361), 
the Guangdong Basic and Applied Basic Research Foundation under Grant No. 2024A1515011949, 
the Shenzhen Science and Technology Program under Grant No. GXWD20231130110308001, JCYJ20250604145617023., and JCYJ20240813104843058,
the Shenzhen Education Science "14th Five-Year Plan" 2023 Annual Project on Artificial Intelligence Special Project under Grant No. rgzn23001, 
the Guangdong Province Higher Education Research and Reform Project under Grant No. YueJiaoGaoHan(2024) No.9 (1227), 
and the Guangdong Province General Colleges and Universities Innovation Team Project under No. 2022KCXTD038. 
\end{acks}

\bibliographystyle{ACM-Reference-Format}
\bibliography{bib}

\appendix

\section{Introduction to Smoothness from a Spectral Perspective}
\label{preliminary: smoothness}

From a spectral perspective, a graph learning task can be viewed as a graph signal processing task, where Graph Neural Networks (GNNs) act as filters. Given a specific homogeneous graph \(\mathcal{G}\) with \(\mathcal{N}\) nodes, denoted as \(\mathcal{V} = \{v_{1}, v_{2}, \dots, v_{\mathcal{N}}\}\), whose edges are represented by an adjacency matrix \(A = [a_{ij}]_{\mathcal{N} \times \mathcal{N}}\), where
\begin{align}
    a_{ij} = \begin{cases}
        1 & \text{if there is an edge between } v_{i} \text{ and } v_{j},\\
        0 & \text{otherwise,}
    \end{cases}
\end{align}
these nodes are characterized by \(d\) features, represented as \(\mathcal{X} = [X_{1}, X_{2}, \dots, X_{d}] \in \mathbb{R}^{\mathcal{N} \times d}\). Each column \(X_{j} = [x_{j1}, x_{j2}, \dots, x_{j\mathcal{N}}]^{T}\), for \(j \in \{1, 2, \dots, d\}\), constitutes a graph signal defined on \(\mathcal{G}\), with \(x_{ji} \in \mathbb{R}\) denoting the \(j\)-th feature value of node \(v_{i}\).

The Laplacian matrix \(L\) is a fundamental element in analyzing graph structures and is defined as:
\begin{align}
    L = D - A,
\end{align}
where \(D\) is the diagonal degree matrix with \(D_{ii} = \sum_{j=1}^{\mathcal{N}} A_{ij}\). Being real and symmetric, \(L\) admits an orthogonal diagonalization:
\begin{align}
    L = \mathcal{U} \Lambda \mathcal{U}^{T} = \mathcal{U} \begin{bmatrix}
        \lambda_{1} & & & \\
        & \lambda_{2} & & \\
        & & \ddots & \\
        & & & \lambda_{\mathcal{N}}
    \end{bmatrix} \mathcal{U}^{T},
\end{align}
where \(\{\lambda_{i}\}_{i=1}^{\mathcal{N}}\) are the eigenvalues of \(L\), satisfying \(\lambda_{1} \leq \lambda_{2} \leq \cdots \leq \lambda_{\mathcal{N}}\), and 
\begin{equation}\label{eq: eigenvector}
    \mathcal{U} = [u_{1}, u_{2}, \dots, u_{\mathcal{N}}]
\end{equation}
is the matrix whose columns are the normalized eigenvectors of \(L\).

Eigenvalues carry rich semantic information, bridging the spectral and spatial domains. Specifically, eigenvectors corresponding to small eigenvalues represent low-frequency components of the graph signal, meaning the signal varies slowly across the graph, which results in neighboring nodes having similar feature values, thus indicating greater smoothness.

Based on these definitions, the general formulation of a GNN is given by:
\begin{align}
    x_{\text{out}} = \mathcal{U} f(\Lambda) \mathcal{U}^{T} x_{\text{in}},
\end{align}
where \(x_{\text{in}}\) and \(x_{\text{out}}\) denote the input and output graph signals, respectively. The function \(f(\Lambda)\) represents the GNN’s frequency response function, serving as the convolutional kernel. By designing different \(f(\Lambda)\), researchers develop various GNN-based models tailored to specific tasks.

For instance, in the Graph Convolutional Network (GCN), a widely used GNN model, the frequency response function is defined as:
\begin{align}
    f(\Lambda) = \theta (I - \Lambda),
\end{align}
which simplifies to \(f(\Lambda) = I - \Lambda\) when \(\theta = 1\). This formulation highlights the low-pass nature of GCNs, as the kernel amplifies low-frequency signals while suppressing high-frequency ones. Since emphasizing low-frequency components in the spectral domain increases the similarity of feature representations between neighboring nodes in the spatial domain, this accounts for the smoothness property observed in GCNs.

\section{Prompts for Generating Descriptions}
\label{appendix: prompt template}

\begin{figure}[htbp]
    \centering
    \includegraphics[width=0.7\textwidth]{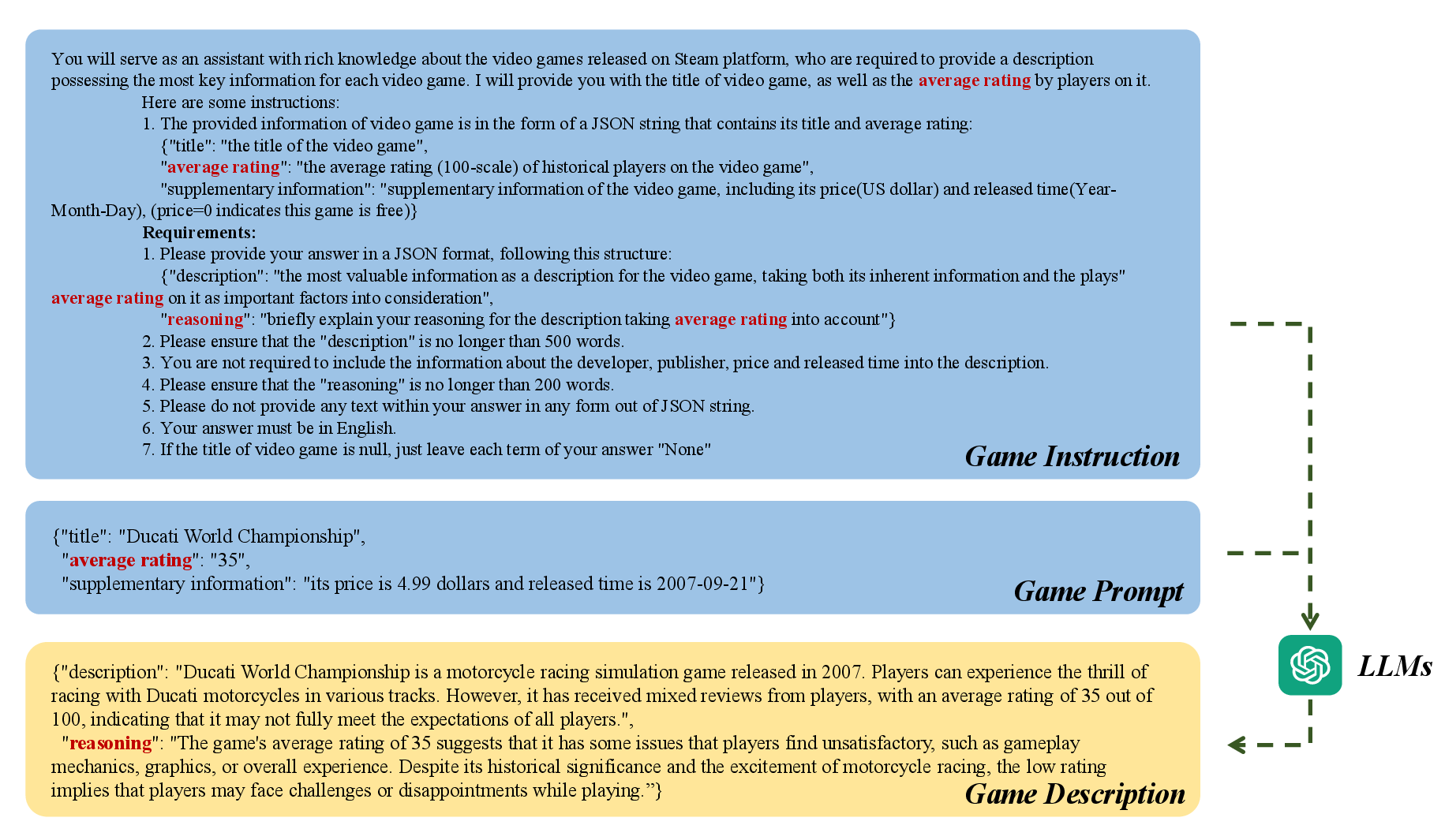}
    \caption{An example of generating a game description.}
\label{fig: prompt template for generating game description}
\end{figure}

\begin{figure}[htbp]
    \centering
    \includegraphics[width=1\textwidth]{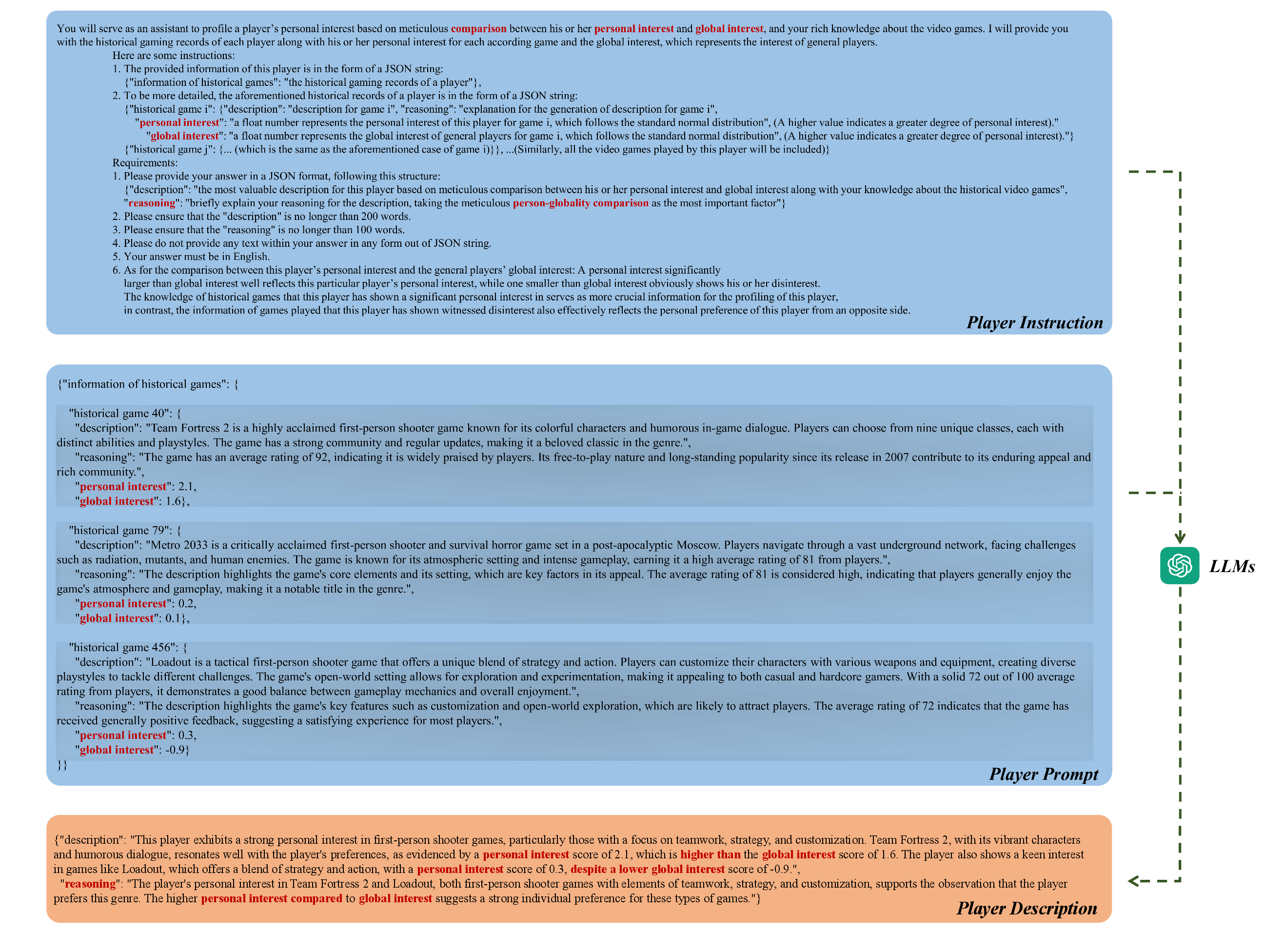}
    \caption{An example of generating a player description.}
\label{fig: prompt template for generating player description}
\end{figure}

This section provides two examples, which show how to generate descriptions for a game and a player, respectively. Specifically, Fig. \ref{fig: prompt template for generating game description} shows the process of generating a description for the game \textit{Ducati World Championship}, whose average rating on the scale 100 is 35. Both \textbf{Game Instruction} and \textbf{Game Prompt} are input to the LLMs, which accordingly outputs the \textbf{Game Description} for \textit{Ducati World Championship}. Fig. \ref{fig: prompt template for generating player description} shows how to generate a description for a player who has played three games whose ID=40, 79, and 456, respectively. Similarly, both \textbf{Player Instruction} and \textbf{Player Prompt} are input to the LLMs, which outputs \textbf{Player Description}. We believe that future work could further refine these prompts using advanced in-context learning strategies~\cite{wqwang1, wqwang2, wqwang3,2026827_1,2026827_2,2026827_3,2026827_4}. When generative components become a serving bottleneck, phase mapping and sparsity-aware truncation provide two relevant directions for reducing inference cost in multimodal generative models~\cite{zhao2026resilphase,zhao2026seeingendstepzero}.

\end{document}